\documentclass[aps,pra,twocolumn]{revtex4-1} 

\usepackage{verbatim}
\usepackage{dsfont} 
\usepackage{amsmath}
\usepackage{amsthm} 
\usepackage{enumerate}
\usepackage{graphicx}
\usepackage{mathtools}
\usepackage{bm}
\usepackage{bbold}
\usepackage{amssymb} 
\usepackage{blindtext}
\usepackage{tcolorbox}
\usepackage{appendix} 
\usepackage{hyperref} 
\usepackage[capitalize,nameinlink]{cleveref}
\hypersetup{
    colorlinks=true,
    linkcolor=green,
    citecolor=magenta,
    unicode=true,  
}

\newtheorem{result}{Result}

\newtheorem{corollary}{Corollary}

\newcommand{\ket}[1]{\left| #1 \right\rangle}

\newcommand{\ketbra}[2]{\left|#1 \rangle \langle #2 \right|}

\DeclareMathOperator{\tr}{Tr}

\usepackage{calc} 
\usepackage{accents}

\usepackage[normalem]{ulem}
\usepackage{textpos}

\begin{document}

\title{Swapping of quantum correlations and the role of local filtering operations}

\author{Pedro Rosario$^{1,5}$}
\email[]{pedrorosario@estudante.ufscar.br}

\author{Andr\'es F. Ducuara$^{2,3,4}$} 
\email[]{andres.ducuara@yukawa.kyoto-u.ac.jp}

\author{Cristian E. Susa$^{5}$}
\email[]{cristiansusa@correo.unicordoba.edu.co}

\address{$^1$Departamento de Física, Universidade Federal de São Carlos, \\Rodovia Washington Luís, km 235 - SP-310, 13565-905 São Carlos, SP, Brazil
\looseness=-1}

\affiliation{$^{2}$Yukawa Institute for Theoretical Physics, Kyoto University, Kitashirakawa Oiwakecho, Sakyo-ku, Kyoto 606-8502, Japan
\looseness=-1}

\affiliation{$^{3}$Center for Gravitational Physics and Quantum Information, Yukawa Institute for Theoretical Physics, Kyoto University
\looseness=-1} 

\affiliation{$^{4}$H.H. Wills Physics Laboratory, University of Bristol, Tyndall Avenue, Bristol, BS8 1TL, United Kingdom 
\looseness=-1}

\address{$^5$Department of Physics and Electronics, University of C\'ordoba, 230002 Monter\'ia, Colombia}





\date{\today}

\begin{abstract}
     We address the swapping of various quantum correlation measures including: Bell-nonlocality, EPR-steering, usefulness for teleportation, entanglement, quantum obesity, as well as the effect that local filtering operations have on the swapping of such correlations. In the first part of this work we address the raw swapping protocol (i. e. without local filtering) and our findings are as follows. First, using the Bloch representation of quantum states, we show that all of the above properties of a general quantum state can fully be preserved whenever the state is swapped together with arbitrary combinations of Bell states and Bell measurements. This generalises a result shown for the concurrence of states in the X-form \cite{s8}. Second, we derive an explicit formula for the quantum obesity of the final post-swapping state in terms of the obesity of general input states and measurements, and therefore establishing the limit at which obesity can be swapped. In the second part we address the effect of local filtering operations on the swapping of quantum correlations. Specifically, we explore whether experimentalists should  implement local filters \emph{before} or \emph{after} the swapping protocol takes place, so in order to maximize the final amount of correlations. In this regard, we first show that these two scenarios are equivalent for the family of Bell-diagonal states, for all of the above-mentioned quantum correlations. We then prove that applying local filters \emph{first} can be more efficient when considering the strictly larger family of almost Bell-diagonal states, with the quantum obesity as the test property. Finally, we provide numerical evidence for this latter phenomenon (local filtering first is more efficient) holding true for general two-qubit states in the X-form, for all of the above-mentioned quantum correlations.
\end{abstract} 
\maketitle

\begin{textblock*}{3cm}(17cm,-13.5cm)
  \footnotesize YITP-23-85
\end{textblock*}
\vspace{-1cm}
\section{Introduction}
\label{s:intro}

Quantum correlations play a fundamental role in the theory of quantum information as well as in the development of quantum technologies \cite{review_correlations, review_di1, review_di2, review_2qr}. The field of quantum communication, in particular, exploits the properties of quantum mechanical systems in order to implement information-theoretic tasks such as the celebrated entanglement swapping protocol \cite{s1,s2}. The original version of this protocol involves labs 1 and 2 sharing a bipartite system in a Bell state (say pair of photons), with lab 2 sharing another Bell state with a third lab. Lab 2 implements a Bell measurement on its pair of systems and, as a consequence of this, labs 1 and 3 end up sharing a Bell state. The remarkable characteristic of this protocol is that physical systems 1 and 4 do not need to have interacted during any point of the process and still, they happen to end up entangled when the protocol finishes. This then leads to the possibility of systems 1 and 4 being arbitrarily spatially separated and so, the protocol naturally leads to long-distance entanglement-based applications opening the door to quantum repeaters, the fundamental block for quantum networks, and a potentially full fledged quantum internet in the long term. In this regard, consistent progress on entanglement swapping has been pushed forward during the past two decades \cite{s3, s4, s5, s6, s7, s6}, with the feasibility of its implementation as quantum repeaters being experimentally verified multiple times \cite{r1, r2, r3, r4, r5, r6, r7, r8, r9}. 

Notwithstanding the promising characteristics of the entanglement swapping protocol, the original version presented above is however a very idealised scenario and so, more realistic versions of this protocol have been explored during the past two decades. Three important aspects to consider here are \cite{r5, r6}: i) the effect of noise on both states and measurements, ii) the swapping of quantum properties beyond entanglement, and iii) additional mechanisms to improve the performance of the protocol. We now address these aspects in more detail.

First, regarding the effect of noise, the original entanglement swapping protocol is a very idealised situation where the shared states are Bell states (maximally entangled pure states) and the measurement is a Bell measurement (maximally entangled projective measurement) \cite{s1,s2}. In more realistic scenarios however, the effect of noise in either the input states or the measurement process would limit the efficiency of the protocol. For instance, whilst the final post-swapping state might still possess a finite amount of entanglement, this might not be the full initial amount, and so entanglement effectively gets degraded. Despite of this however, one can still think about these \emph{noisy} swapping scenarios as effectively \emph{extending} the reach of entanglement, at the price of not fully preserving its initial amount {\cite{Bej_2022}}. The interplay between the amount of entanglement being swapped, noise, as well as the protocol's success probability is an active subject of research \cite{f9, f10, f11, kirby_2020, kirby_2021}. 

A second aspect to consider is that, whilst the swapping protocol was first introduced in the context of entanglement, the protocol can similarly be explored for additional quantum properties and so, in more general terms, we can talk about the \emph{swapping of quantum correlations}. In this regard, whilst some additional properties have been explored within swapping scenarios including Bell's nonlocality \cite{s3,s5,s6} as well as usefulness for teleportation \cite{s4}, many other quantum properties of interest remain currently unexplored.

Third, it is of foundational interest as well as practical importance to develop mechanisms so to try to preserve/enhance the amount of correlations that are being swapped. In this regard, there exist mechanisms like distillation, multi-copy scenarios, local filtering operations, amongst others \cite{r5, r6}. Local filtering operations \cite{f1, f2, f3, f4, f5, f6, f7, f8, f12, f13, f14, f15}, or stochastic local operations and classical communication (SLOCC) more formally, are experimentally-friendly transformations which can be implemented with the aid of local measurements and classical communication \cite{fe1, fe2, fe3, fe5, fe6}. The effect of local filtering on the swapping protocol, to the best of our knowledge, has not been addressed before and so, our main motivation here is to start exploring this direction of research. In particular, taking into account that the involved parties are interested in maximising the amount of correlations to be swapped, a natural question that emerges is whether they should implement local filtering operations \emph{before} or \emph{after} the swapping protocol takes place. This is one of the main questions we aim to explore in this manuscript.

In this work we address the swapping of various quantum correlations whilst integrating the three previous general points as follows. First, we consider general swapping scenarios with arbitrary quantum input states (not restricted to maximally correlated states), and general POVM measurements (not restricted to projective measurements). Second, we address various quantum correlation measures beyond entanglement \cite{review_entanglement} which are of general interest for the development of quantum technologies: Bell-nonlocality \cite{review_nonlocality}, EPR-steering \cite{review_steering1, review_steering2}, usefulness for teleportation \cite{review_teleportation}, and quantum obesity \cite{Milne_2014_2}. Third, we consider swapping scenarios under local filtering operations, and compare the performance of two natural procedures, one where the local filtering is implemented \emph{before} the swapping protocol, and the other way around, a procedure where the local filtering is implemented \emph{after} the swapping takes place. Within this framework, our results are as follows.

In the first part of this work we address the raw swapping protocol (i. e. without local filtering) and show the following. First, we prove that the quantum correlations of \emph{general} input states can \emph{fully} be preserved when the state is swapped together with arbitrary Bell states and arbitrary Bell measurements. This analytical construction clarifies numerical evidence previously found in the literature \cite{s8}, and it also generalises an analytical result previously found for the specific case of the concurrence of two-qubit states in the X-form \cite{s8}. Specifically, the result we provide here holds for \emph{general} input states (X-form states included), as well as for \emph{general} correlation measures which are invariant under local unitaries (concurrence included). Second, we derive an explicit formula for the quantum obesity of the post-swapping state, in terms of the obesities of general input states and measurements, and therefore explicitly establishing the amount of obesity which can be extracted under swapping protocols. 

In the second part of this work we explore the effect of local filtering operations on the swapping protocol by considering two natural scenarios: i) implementing filtering first and then swapping (FS), and the other way around ii) implementing swapping first and then filtering (SF). We first show that these two protocols are equivalent for Bell-diagonal states, for all the considered quantum correlations. We then show that this equivalence breaks down when considering strictly larger families of states. Specifically, we show that the FS protocol is strictly better than the SF protocol for the strictly larger class of almost-Bell-diagonal states, using the quantum obesity as the test property. Finally, we provide numerical evidence for the relationship ``$\rm FS\geq SF$" holding true for arbitrary two-qubit states in the X-form. 
 
This work is organised as follows. In Section \ref{sec:background} we address the background theory: the families of quantum states relevant to our findings, the description of various quantum correlation quantifiers, the swapping protocol, and local filtering operations. In Section \ref{sec:main_result} we present our main results: In Sec.~\ref{sec:general_swapping} we derive the final post-swapping state in the Bloch representation for general swapping scenarios (including qudit and multi-qubit cases). In Sec.~\ref{sec:FS_SF_interplay} we derive closed formulae for the quantum obesity of the final post-swapping state and discuss the interplay between local filtering and the swapping of quantum correlations. Finally, we present some conclusions in Section \ref{sec:conclusions}.

\section{Background theory} 
\label{sec:background}

In this section we address the preliminaries needed to establish our main results. We address families of states of interest, quantum correlation measures, the swapping protocol, and local filtering operations.  

\begin{figure}[h!]
\centering
\includegraphics[scale=1]{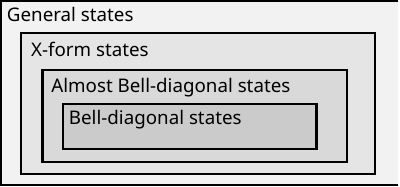}
\vspace{-0.2cm}
\caption{Hierarchy of sets of quantum states discussed in this work. General states $\supseteq$ X-form states $\supseteq$ Almost Bell-diagonal (ABD) states  $\supseteq$ Bell-diagonal (BD) states $\supseteq$ Bell states.}
\label{fig:states_set}
\end{figure}

\emph{Hierarchy of sets of states.} The families of states of interest in this work are: general states, X-form states, almost Bell-diagonal states, Bell-diagonal states, and Bell states. These families define the hierarchy depicted in \autoref{fig:states_set} and we describe them as follows. A general two-qubit state $\rho \in D(\mathds{C}^2 \otimes \mathds{C}^2)$ is a positive semidefinite ($\rho \geq 0$) trace one ($\tr (\rho)=1$) operator on $\mathds{C}^2 \otimes \mathds{C}^2$. These states can be written as:
\begin{align}
    \rho
    =
    \frac{1}{4}
    \sum^{3}_{i,j=0}
    R_{ij}
    \sigma_{i}
    \otimes 
    \sigma_{j}
    ,
    \hspace{0.5cm}
    R
    =
    \begin{pmatrix}
    1 & \vec{b}^{T}
    \\
    \vec{a} & T 
    \end{pmatrix}
    ,
    \label{eq:R_matrix}
\end{align}
with $\{\sigma_{1}, \sigma_{2}, \sigma_{3}\}$ the Pauli matrices, $\sigma_{0} = \mathbb{1}$, local vectors $\vec{a}=(a_1, a_2, a_3)$, with $a_{i} = \tr [ (\sigma_{i} \otimes \mathbb{1}) \rho ]$, $\vec{b}=(b_1, b_2, b_3)$, with $b_{i} = \tr [ (\mathbb{1} \otimes \sigma_{i}) \rho ]$, and the correlation matrix $T_{ij} = \tr [ \rho( \sigma_{i} \otimes \sigma_{j}) ]$ {\cite{omar_2016}}.  As elements in \eqref{eq:R_matrix} identify the quantum state $\rho$, we refer to the former as the R-representation or R-picture of the state $\rho$. Alternatively, we can also say that a general two-qubit state is characterised by the triple $(\vec{a}, \vec{b}, T)$. A \emph{X-form} state is a general state whose only non-zero elements are in the diagonal and anti-diagonal, hence their name. These states satisfy the condition $\vec{a}=(0, 0, a_3)$ and $\vec{b}=(0, 0, b_3)$. An \emph{almost Bell-diagonal} (ABD) state is an X-form state with the additional constraint $a_3=\pm b_3$. A \emph{Bell-diagonal} (BD) state is an almost Bell-diagonal state with the extra condition $a_3=b_3=0$. Finally, the four Bell states can be written as 
$
\Phi_{n} 
=
(
\mathds{1}
\otimes
\sigma_n
)
\Phi^-
(
\mathds{1}
\otimes
\sigma_n
)
$, 
$n=0,1,2,3$,
$
\Phi^-
\coloneqq
\ketbra{\Phi^-}{\Phi^-}
$,
$
\ket{\Phi^-}
\coloneqq
(1/\sqrt{2})(\ket{01}-\ket{10})
$. In particular, the R-representation of the singlet $\Phi^{-}$ reads $R=\text{diag}(1,-\mathds{1})$. We also consider two-qudit quantum states and write them as:
\begin{align}
    \boldsymbol{\rho}
    =
    \frac{1}{d^{2}}
    \sum_{i,j=0}^{d^{2}-1}
    \mathbf{R}_{ij}
    \boldsymbol{\sigma}_{i}
    \otimes 
    \boldsymbol{\sigma}_{j}
    ,
    \label{eq:R_matrixd}
\end{align}
with $\boldsymbol{\sigma}_{0} = \mathbb{1}_{d}$, $\{ \boldsymbol{ \sigma_{i}} \}_{i=1}^{d^{2}-1}$ a set of ($d\times d$)-operators satisfying
$
\tr
[
\boldsymbol{
\sigma
}_{i}
\boldsymbol{
\sigma
}_{j}
]
=
d
\delta_{ij}
$,
$\tr\hspace{2pt}
\left[
\boldsymbol{\sigma}_{i}
\right]=0
$
,
$
i,j=0,1,...,d^{2}-1
$,
$\mathbf{R}_{ij} = \tr
[
(
\boldsymbol{\sigma}_{i}
\otimes 
\boldsymbol{\sigma}_{j}
)
\boldsymbol{\rho}
]$
{\cite{Jing_2022_1,Br_ning_2012}}.
We now address quantum measurements, or positive operator-valued measures (POVM). A general two-qudit POVM $\{E_k\}$ is a set of  positive semidefinite operators ($E_k \geq 0$, $\forall k$)  satisfying $\sum_k E_k = \mathbb{1}$. The operators $E_k$ are addressed as POVM elements and they each can be written as:
\begin{align}
  \boldsymbol{E}
  =
  \sum_{k,l=0}^{d^{2}-1}
  \boldsymbol{
  \mathfrak{R}
  }_{kl}
  \boldsymbol{\sigma}_{k}
  \otimes
  \boldsymbol{\sigma}_{l}
  ,
  \label{eq:POVM_element}
\end{align}
$
\boldsymbol{
\mathfrak{R}
  }_{kl}
  =
  \tr
  [
(
\boldsymbol{\sigma}_{k}
\otimes 
\boldsymbol{\sigma}_{l}
)
\boldsymbol{E}
]
$.
As elements in \eqref{eq:POVM_element} identify the POVM effect $E$, we say that the POVM element is represented by its 
$
\boldsymbol{
\mathfrak{R}
}
$
matrix. Similarly to the case for qubits, we refer to two-qudit states and POVM elements  satisfying $\tr [ (\boldsymbol{\sigma}_{i} \otimes \mathbb{1}) \boldsymbol{O} ]=0$, and $\tr [ (\mathbb{1} \otimes \boldsymbol{\sigma}_{i}) \boldsymbol{O} ]=0$, $\forall i$, $\boldsymbol{O}\in \{\boldsymbol{\rho}, \boldsymbol{E}\}$ as \emph{Bell-diagonal}.

\begin{figure}[h!]
\centering
\includegraphics[scale=1]{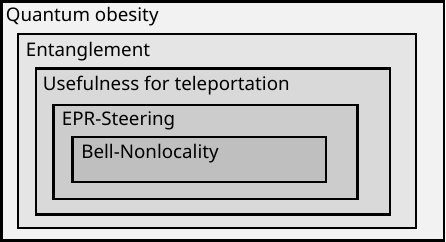}
\vspace{-0.2cm}
\caption{Hierarchy of quantum correlations of interest in this work. All nonlocal states are steerable but some steerable states are not nonlocal. All steerable states are useful for teleportation (UFT) but some UFT states are not steerable. All UFT states are entangled but some entangled states are not useful for teleportation. All entangled states are obese but some obese states are separable.}
\label{fig:correlations}
\end{figure}

\emph{Quantum correlation measures.} We now define correlation measures for the quantum properties of interest: Bell-nonlocality, EPR-steering, usefulness for teleportation, entanglement, and quantum obesity. These quantum properties define the hierarchy depicted in \autoref{fig:correlations}, and we introduce them as follows. As a measure for Bell-nonlocality we consider the violation of the CHSH-inequality. In this regard, the Horodecki criterion establishes that a given state $\rho$ can violate the CHSH-inequality if and only if $s^{2}_{1} + s^{2}_{2} > 1$ {\cite{Horodecki_1995,ducuara_2020}}, with $s_{1},s_{2},s_{3}$ the singular values of $T$ in decreasing order $s_{1} \geq s_{2} \geq s_{3}$. With this criterion in place, a CHSH-nonlocality correlation measure can naturally be defined as:
\begin{align}
 \mathrm{B}
 (\rho)
 =
 \max 
 \left\{
 0
 ,
 s^{2}_{1} + s^{2}_{2} - 1
 \right\}.
 \label{eq:chsh_measure}
\end{align}
As a measure for EPR-steering we consider the violation of the $\mathrm{F}_{3}$-inequality. The Costa-Angelo criterion establishes that a given two-qubit state can violate the $\mathrm{F}_{3}$-inequality if and only if $s^{2}_{1}+s^{2}_{2} +s^{2}_{3}>1$ \cite{costa_2016}. A correlation measure for $\mathrm{F}_{3}$-inequality violation can then be defined as:
\begin{align}
    \mathrm{BF}_{3}
    (\rho)
    =
    \max 
    \left\{0,
    (1/2)
    (
    s^{2}_{1}+s^{2}_{2}+s^{2}_{3} -1
    )
    \right\}
    ,
    \label{eq:steering_measure}
\end{align}
with the $1/2$ being introduced to guarantee normalisation ($\mathrm{BF}_{3} \leq 1$). A quantum state is useful for
teleportation (UFT) if and only if $|s_{1}| + |s_{2}|-\chi|s_{3}|>1$ {\cite{Horodecki_1996}}, with $\chi=\det T$ the so-called chirality \cite{Milne_2014, Milne_2014_2}. Similarly to Bell-nonlocality and EPR-steering, a measure for UFT can then be defined as:
\begin{align}
    \mathrm{D}(\rho)
    =
    \max
    \left\{
    0
    ,
    (1/2)
    (
    |s_{1}|+|s_{2}|-\chi|s_{3}|-1
    )
    \right\}
    ,
    \label{eq:UFT_measure}
\end{align}
the $1/2$ is again introduced to guarantee normalisation ($\mathrm{D} \leq 1$). Quantum correlation measures quantified by \eqref{eq:chsh_measure}, \eqref{eq:steering_measure} and \eqref{eq:UFT_measure} are entanglement-dependent resources, meaning that they achieve zero for separable states. We quantify entanglement by means of the concurrence:
\begin{align}
    \mathrm{C}(\rho)
    =
    \max
    \left\{
    0
    ,
    \sqrt{\lambda_1}
    -
    \sqrt{\lambda_2}
    -
    \sqrt{\lambda_3}
    -
    \sqrt{\lambda_4}
    \right\}
\end{align}
where $\lambda_1> \lambda_2 > \lambda_3 > \lambda_4$ are the eigenvalues of  $\rho(\sigma_{2}\otimes\sigma_{2})\rho^{*}(\sigma_{2}\otimes \sigma_{2})$, with $\rho^*$ the complex conjugate of the density operator \cite{Wootters_1998}. The final property of interest in this work is the so-called quantum obesity \cite{Milne_2014, Milne_2014_2}. The \emph{quantum obesity} of a two-qudit state reads:
\begin{align}
    \Omega(
    \boldsymbol{\rho}
    )
    =
    |
    \det \mathbf{R}
    |^{1/d^2}
    ,
    \label{eq:obesity}
\end{align}
with the matrix $\mathbf{R}$ as in \eqref{eq:R_matrixd}. This quantum property receives its name due to being related to the \emph{volume} of the quantum steering ellipsoids associated to the quantum state in question \cite{Bhosale_2016, Milne_2014_2}. For two-qubit states, the quantum obesity \eqref{eq:obesity} is normalised ($\Omega (\rho) \in [0,1]$, $\forall \rho$) and an upper bound for the concurrence ($\mathrm{C}(\rho) \leq \Omega (\rho)$, $\forall \rho$) \cite{Bhosale_2016, Milne_2014_2}.  

\begin{figure}[h!]
\centering
\includegraphics[scale=0.6]{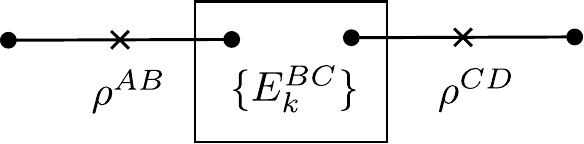}
\vspace{-0.2cm}
\caption{{\bf Swapping protocol}: Pairs of particles $(\mathrm{A},\mathrm{B})$ and $(\mathrm{C},\mathrm{D})$ come from different sources. A bipartite measurement $\{E_k^{\rm BC}\}$ is implemented on particles $(\mathrm{B},\mathrm{C})$. The swapping protocol explores the properties of the final state $\rho^{\rm AD}_k$ \eqref{eq:state_swap} for particles $(\mathrm{A},\mathrm{D})$ after the measurement takes place.}
\label{fig:swapping}
\end{figure}

\emph{General swapping protocol.} Consider a four-partite quantum state of the form $\rho^{AB} \otimes \rho^{CD}$ representing a pair of particles $(\mathrm{A},\mathrm{B})$ in state $\rho^{\rm AB}$, and a pair of particles $(\mathrm{C},\mathrm{D})$ in $\rho^{\rm CD}$ being prepared from independent sources. Consider also that particles $B$ and $C$ are sent to the same lab in which a bipartite measurement $\{E_k^{BC}\}$ is implemented, while particles $A$ and $D$ are sent far apart to each other. In \autoref{fig:swapping} we illustrate the elements of this general swapping protocol. The final state of interest, or post-swapping state $\rho^{AD}$, can be determined  by the generalised von Neumann-L\"{u}ders transformation rule \cite{Barnum_2002,s8}:
\begin{align}
    \rho_{k}^{\mathrm{AD}} 
    &\equiv    
    \frac{
    \tr_{\mathrm{BC}}
    \left[
    (\rho^{AB} \otimes \rho^{CD})
    (
    \mathbb{1}^{A}
    \otimes 
    E_k^{\mathrm{BC}}
    \otimes
    \mathbb{1}^{D}
    )
    \right]
    }{
    p_k
    },
    \label{eq:state_swap}
\end{align}
where the outcome probability reads 
$
p_k
=
\tr[
(\rho^{AB} \otimes \rho^{CD})
(
\mathbb{1}^{A}
\otimes 
E_k^{\mathrm{BC}}
\otimes
\mathbb{1}^{D}
)
]
$. 
In particular, when $\rho^{AB}$ and $\rho^{CD}$ are Bell states and the POVM is a Bell measurement, the post-swapping state $\rho^{AD}$ turns out also to be a Bell state. This means that the final state $\rho^{AD}$ displays (maximal) quantum correlations, even though particles $A$ and $D$ do not need to have interacted during any point in the past, and so the correlations of the input states have effectively been `` swapped".

\begin{figure}[h!]
\centering
\includegraphics[scale=0.6]{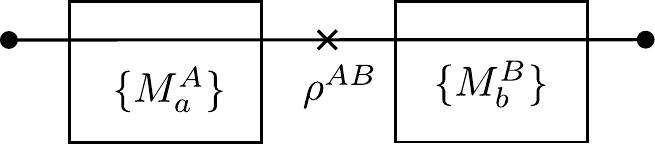}
\vspace{-0.2cm}
\caption{{\bf Local filtering}: A target local filtering operation, say ($f^{\rm A}, f^{\rm B}$), can be implemented by considering local POVMs containing at least a POVM element given according to the desired filters as $M^{\rm A} \coloneqq (f^{\rm A})^{\dagger} f^{\rm A}$, $M^{\rm B} \coloneqq (f^{\rm B})^{\dagger} f^{\rm B}$. In particular, for a general two-qubit state $\rho^{\rm AB}$, we consider here the the KLM-SLOCC local filtering operation taking $\rho^{\rm AB}$ to its Bell-diagonal form. 
}
\label{fig:filtering}
\end{figure}

\emph{Local filtering operations.} Local filtering operations, or stochastic local operations and classical communication (SLOCC) more formally, are operations which can be implemented by means of local measurements and rounds of classical communication. In \autoref{fig:filtering} we illustrate the implementation of local filtering operations and we describe them as follows. Consider a bipartite quantum state $\rho^{\rm AB}$ shared between parties A and B. Allowing the parties to implement local measurements amounts to transforming the state to unnormalised states of the form
$
(
f^{A}
\otimes 
f^B
)
\rho^{\rm AB}
(
f^{A}
\otimes 
f^B
)^{\dagger}
$. Allowing now the parties to communicate their local measurement outcomes as well as to repeat the local measurement process until arriving to a desired final state leads to general local filtering operations which can be written as:
\begin{align}
    (\rho^{\rm AB})'
    =
    \frac{(f^{\mathrm{A}}
    \otimes f^{\mathrm{B}})
    \hspace{2pt}
    \rho^{\rm AB}
    \hspace{2pt}
    (
    f^{\mathrm{A}}
    \otimes 
    f^{\mathrm{B}}
    )^{\dagger}}
    {\tr\hspace{2pt}
    [(f^{\mathrm{A}}
    \otimes f^{\mathrm{B}})
    \hspace{2pt}
    \rho^{\rm AB}
    \hspace{2pt}(f^{\mathrm{A}}
    \otimes f^{\mathrm{B}})^{\dagger}]}.
    \label{eq:locc_state}
\end{align}
with $f^{\mathrm{A}}$, $f^{\mathrm{B}}$ arbitrary positive semidefinite operators satisfying $(f^{\rm A})^{\dagger}f^{\rm A} \leq \mathbb{1}$, $(f^{\rm B})^{\dagger} f^{\rm B} \leq \mathbb{1}$. The operators $f^{\mathrm{A}}$ and $f^{\mathrm{B}}$ are addressed as \emph{local filters}, and are assumed to be invertible since non-invertible maps always decrease entanglement \cite{f6}. Among all possible local filtering operations we address here the local filters that map a general state $\rho$ into its so-called Bell-diagonal form \cite{f5}. We address this map as the Kent-Linden-Massar SLOCC (KLM-SLOCC) of a given state $\rho$ \cite{f5}. The KLM-SLOCC is a map acting as $\rho \rightarrow \rho_{\mathrm{BD}}$, or equivalently as  $\mathrm{R}\rightarrow \mathrm{R}_{\mathrm{BD}} = \text{diag}(1,\sqrt{\nu_{1}/\nu_{0}},\sqrt{\nu_{2}/\nu_{0}},-\sqrt{\nu_{3}/\nu_{0}})$ with $\{\nu_{0}, \nu_{1}, \nu_{2}, \nu_{3}\}$ the eigenvalues of $\eta R \eta R^{T}$ in decreasing order $\nu_{0}\geq \nu_{1}\geq \nu_{2} \geq \nu_{3}$ with $\eta = \text{diag}(1,-1,-1,-1)$ {\cite{pal_2015}}. The KLM-SLOCC has been proven to simultaneously maximise the violation of CHSH-inequality {\cite{f8}}, usefulness for teleportation {\cite{f7}}, and the concurrence {\cite{f6}}. 

\section{Main results}
\label{sec:main_result}
We present our main results as follows. In Section \ref{sec:general_swapping}, we derived a transformation rule for general swapping scenarios in the $R$-representation of quantum states. This result is then extended to two-qudit as well as multi-source swapping scenarios. In Section \ref{sec:FS_SF_interplay}, we discuss the joint action of the swapping protocol and local filtering operations on quantum correlations.

\subsection{Swapping protocol in the Bloch representation and the swapping of quantum obesity}
\label{sec:general_swapping}

In this subsection we derive a result regarding the nature of the $R$-representation of the post-swapping state in terms of the $R$-representation of the initial states and measurements. We first derive this relationship for general input states and arbitrary input POVM element for two-qubit systems. We then analyse the particular case when the state of (C,D) is an arbitrary Bell state and the measurement is an arbitrary Bell POVM element. We then address a generalisation to two-qudit systems as well as a chain of $N$ qubits.

\begin{result}[General swapping protocol in the Bloch representation]
\label{r:general_swapping}
Let $\rho^{\mathrm{AB}}$ and $\rho^{\mathrm{CD}}$  be two general two-qubit states characterised by $R^{\mathrm{AB}}$ and $R^{\mathrm{CD}}$ \eqref{eq:R_matrix}, and $E^{\rm  BC}$ be a general two-qubit POVM element characterised by $\mathfrak{R}^{\mathrm{BC}}$. The post-swapping state $\rho^{\mathrm{AD}}$ is then represented by a matrix $R^{\mathrm{AD}}$ that takes the form:
\begin{align}
    R^{\mathrm{AD}}
    =
    \frac{
    R^{\mathrm{AB}}
    \mathfrak{R}^{\mathrm{BC}}
    R^{\mathrm{CD}}
    }
    {
    \left[
        R^{\mathrm{AB}}
        \mathfrak{R}^{\mathrm{BC}}
        R^{\mathrm{CD}}
    \right]_{00}
    }
    ,
    \label{eq:r1}
\end{align}
with $[X]_{00}$ the (0,0) component of the matrix $X$.
\end{result}
The proof of this result is given in \cref{app:proof_general_swapping}. 
This result explicitly provides the R-representation of the post-swapping state $\rho^{\rm AD}$ in terms of the R-representation of the initial states and the POVM element. In particular, we can use this to readily recover the standard swapping protocol as follows. Take both input states and the POVM element to be the singlet, meaning that their R-representation reads $R^{\rm AB} = R^{\rm CD}= \mathfrak{R}^{\rm BC}= \text{diag}(1,-\mathds{1})$. We then have
$[
    R^{\mathrm{AB}}
    \mathfrak{R}^{\mathrm{BC}}
    R^{\mathrm{CD}}
]_{00}
=
1
$ and, using Result \ref{r:general_swapping}, we get that the R-representation of the final state reads $R^{\rm AD} = R^{\rm AB} \mathfrak{R^{\rm BC}} R^{\rm CD}=\text{diag}(1,-\mathds{1})$. This means that the final system is in the singlet state and so, recovering the statement of the original swapping protocol. We can similarly leave $R^{\rm AB}$ representing a general state and instead only impose $R^{\rm CD} = \mathfrak{R}^{\rm BC} = \text{diag}(1,-\mathds{1})$, from which we now get $R^{\rm AD} = R^{\rm AB} \mathfrak{R}^{\rm BC} R^{\rm CD} = R^{\rm AB}$, meaning that the input state remains invariant. This latter argument does not depend on the specific Bell state being used, since the R-representation of any Bell state $\Phi_n$ satisfies $R^2=\mathds{1}$. Finally, following this line of thought, we can also ask about the behaviour of the final state when considering a general input state $\rho^{\rm AB}$, together with arbitrary combinations of Bell states in CD $\rho^{\rm CD}=\Phi_n^{\rm CD}$ and arbitrary Bell POVM elements in BC $E^{\rm BC}=\Phi_m^{\rm BC}$, $n,m=0,1,2,3$. Within this specific scenario, it has analytically been proven that the concurrence of the final state coincides with that of the input state, for input states in the X-form \cite{s8}. An analytical construction for this statement holding true for general two-qubit input states was left as an open question in \cite{s8}, and it was supported by numerical evidence. We now provide a proof for that statement using the R-representation of quantum states an measurements.

\begin{corollary}[Swapping protocol with general input state $\rho^{\rm AB}$ and arbitrary combinations of Bell states and POVM elements]
\label{cr:standard_swapping}
Let $\rho^{\mathrm{AB}}$ be a general two-qubit state represented by $R^{\rm AB}$ \eqref{eq:R_matrix}, $\rho^{\rm CD} \coloneqq \Phi_n^{\rm CD}$ any Bell state $n=0,1,2,3$,  
$E^{\rm BC} 
\coloneqq 
\Phi_m^{\rm BC}
$, any Bell projector $m=0,1,2,3$. The post-swapping state then reads: 
{\small\begin{align}
    \rho^{\mathrm{AD}}_{n,m}
    =
    \left(
    \mathbb{1}
    \otimes 
    \sigma_{n}\sigma_{m}
    \right)
    (
    \frac{1}{4}\sum^{3}_{i,j=0}
    \mathrm{R}^{\mathrm{AB}}_{ij}
    \sigma^{\mathrm{A}}_{i}
    \otimes \sigma^{\mathrm{D}}_{j}
    )
    \left(
    \mathbb{1}
    \otimes 
    \sigma_{m}\sigma_{n}
    \right)
    .
    \label{eq:proposition1}
\end{align}}
\end{corollary}
Details about the derivation of this statement is in \cref{app:proof_standard_swapping}. This corollary tells us that any of the eight possible combinations of post-swapping states $\rho^{\mathrm{AD}}_{n,m}$ are identical, up to the action of local unitary operations $\sigma_{m}\sigma_{n}$, to the input state $\rho^{\mathrm{AB}}$. In particular, this means that both the initial and final states display the same type of correlations, for all quantum correlation measures which are invariant under local unitary operations. In particular, this then holds for all the correlation measures considered in this work, the concurrence in particular. This analytical construction then effectively generalises that derived in \cite{s8} for the concurrence of states in the X-form. We now move on to provide another application of result 1, and derive a closed formula for the quantum obesity of the final post-swapping state for general input states and POVM elements. 
\begin{corollary}[General swapping of quantum obesity]
    Consider general quantum states $\rho^{\rm AB}$, $\rho^{\rm CD}$ represented by $R^{\rm AB}$, $R^{\rm CD}$, and a POVM element $E^{\rm BC}$ represented by $\mathfrak{R}^{\rm BC}$. The quantum obesity of the post-swapping state $\rho^{\mathrm{AD}}$ is given by:
    \begin{align}
    \Omega^{\mathrm{AD}}
    =
    \frac{
    \Omega^{\mathrm{AB}}
    \zeta^{\mathrm{BC}}
    \Omega^{\mathrm{CD}}
    }{
    |
    \left[
        R^{\mathrm{AB}}
        \mathfrak{R}^{\mathrm{BC}}
        R^{\mathrm{CD}}
    \right]_{00}
    |
    }
    ,
    \label{eq:caso4_result4}
    \end{align}
    with  $\Omega^{\mathrm{AB}}$ ($\Omega^{\mathrm{CD}}$) the quantum obesity of the state $\rho^{\mathrm{AB}}$  ($\rho^{\mathrm{CD}}$), $\zeta^{\mathrm{BC}} \coloneqq |\det \mathfrak{R}^{\mathrm{BC}} |^{1/4}$, and $[X]_{00}$ the (0,0) component of the matrix $X$.
    \label{r:r4}
\end{corollary}
This corollary provides an explicit expression for the quantum obesity of the post-swapping state ($\Omega^{\mathrm{AD}}$) in terms of the quantum obesity of the two initial states ($\Omega^{\mathrm{AB}}, \Omega^{\mathrm{CD}}$), as well as from an obesity-like quantity of the POVM element involved in the intermediate measurement process $\zeta^{\mathrm{BC}}$. In particular, if the POVM element is any Bell state, we then get $\zeta^{\rm BC}=1$, and so the relation for obesity reads 
$
\Omega^{\mathrm{AD}}
=
\Omega^{\mathrm{AB}}
\Omega^{\mathrm{CD}}
/c
$,
$c
=
|
[
    R^{\mathrm{AB}}
    \mathfrak{R}^{\mathrm{BC}}
    R^{\mathrm{CD}}
]_{00}
|
$. In this regard, it is worth mentioning that a similar relationship has been conjectured to hold true for the concurrence of general states under the swapping protocol as $C^{\mathrm{AD}} \leq C^{\mathrm{AB}} C^{\mathrm{CD}}$ \cite{s8,Bergou_2021}. This is then telling us that despite representing different quantum properties, entanglement and obesity are governed by similar-looking inequalities. Coming back to the construction in Result \ref{r:general_swapping}, we can now also check that this is not limited for two-qubit states, but can also be extended to higher dimensions as follows.

\begin{result}[Swapping scenario with two two-qudit sources]
\label{r:qudit_swapping}
 Let $\boldsymbol{\rho}^{\mathrm{AB}}$ and $\boldsymbol{\rho}^{\mathrm{CD}}$  two general two-qudit states in $ D(\mathds{C}^d \otimes \mathds{C}^d)$ characterised by $\boldsymbol{\mathrm{R}}^{\mathrm{AB}}$ and $\boldsymbol{\mathrm{R}}^{\mathrm{CD}}$. Let $\boldsymbol{E}^{\rm BC}$ be a general two-qudit POVM element characterized by  $\boldsymbol{\mathfrak{R}}^{\mathrm{BC}}$. The post-swapping state $\boldsymbol{\rho}^{\mathrm{AD}}$ is represented by a matrix
$\boldsymbol{\mathrm{R}}^{\mathrm{AD}}$
 that takes the form:
\begin{align}
\boldsymbol{\mathrm{R}}^{\mathrm{AD}}
=
\frac{
    \boldsymbol{\mathrm{R}}^{\mathrm{AB}}
    \boldsymbol{\mathfrak{R}}^{\mathrm{BC}}
    \boldsymbol{\mathrm{R}}^{\mathrm{CD}}
}
{
\left[ 
    \boldsymbol{\mathrm{R}}^{\mathrm{AB}}
    \boldsymbol{\mathfrak{R}}^{\mathrm{BC}}
    \boldsymbol{\mathrm{R}}^{\mathrm{CD}}
\right]_{00}
}
,
\label{eq:r3}
\end{align}
with $[X]_{00}$ the (0,0) component of the matrix X, $\boldsymbol{\sigma}_{i}$ hermitian operators in $su(d)$  satisfying $\tr[\boldsymbol{\sigma}_{i}]=0$, $\tr[\boldsymbol{\sigma}_{i}\boldsymbol{\sigma}_{j}]=d\delta_{ij}$ for $i,j=1,\dots\,d^2-1$, and $\boldsymbol{\sigma}_{0}= \mathbb{1}_{d}$.
\end{result}
The proof of this result builds on the proof for the two-qubit case and, for completeness, it is addressed in  \cref{app:proof_qudit_swapping}. We now move on to a swapping scenario with multiple sources and measurements. 

\begin{result}[Swapping scenario with $N$ two-qubit sources] 
\label{r:multiq_swapping}
Let the total initial state be composed by $N\geq 2$ two-qubit sources and $N-1$ bipartite measurements as
$
\rho^{p_1...p_{2N}}
\equiv
\rho^{\rm I}
$, and
$
M^{p_2...p_{2N-1}}
\equiv
M^{\rm I}
$
given by:
\begin{align}
    \rho^{\rm I}
    \coloneqq
    \bigotimes_{i=1}^{N}
    \rho^{p_{2i-1},p_{2i}}
    , \hspace{0.6cm}
    M^{\rm I}
    \coloneqq
    \bigotimes_{j=1}^{N-1}
     M^{p_{2j},p_{2j+1}}
     .
     \label{eq:multiq_measure_chain}
\end{align}
The post-swapping state $\rho^{p_1p_{2N}}$ of particles $p_{1}$ and $p_{2\mathrm{N}}$ is represented by a matrix $\mathrm{R}^{p_{1}p_{2\mathrm{N}}}$ that takes the form:
\begin{align}
    \mathrm{R}^{p_{1}p_{2\mathrm{N}}}
    &=
    \frac{
    \prod_{i=1}^{N}\mathrm{R}^{p_{2i-1},p_{2i}}\mathfrak{R}^{p_{2i},p_{2i+1}}
    }
    {
    \left[
    \prod_{i=1}^{N}\mathrm{R}^{p_{2i-1},p_{2i}}\mathfrak{R}^{p_{2i},p_{2i+1}}
    \right]_{00}
    }
    ,
    \label{eq:p1p2N_R}
\end{align}
with  $\mathfrak{R}^{p_{2N},p_{2N+1}}\coloneqq\mathds{1}^{p_{2N},p_{2N+1}}$, and  $[X]_{00}$ the (0,0) component of the matrix $X$.
\end{result}
The proof of this result is in \cref{sec:chainqubits}. We can check that setting $N=2$ recovers \cref{r:general_swapping}. We can also address the behaviour of quantum obesity for the two-qudit case as well as the multi-qubit case as follows.
\begin{corollary}[Quantum obesity for two-qudit and multi-qubit general swapping protocols]
\textbf{a)} The quantum obesity of the post-swapping two-qudit state in \eqref{eq:r3} is given by:
\begin{align}
    \nonumber 
    \boldsymbol{\Omega}^{\mathrm{AD}}
    =
    \frac{
    \boldsymbol{\Omega}^{\mathrm{AB}}
    \boldsymbol{\zeta}^{\mathrm{BC}}
    \boldsymbol{\Omega}^{\mathrm{CD}}
    }{
    \left|
    \left[ 
    \boldsymbol{\mathrm{R}}^{\mathrm{AB}}
    \boldsymbol{\mathfrak{R}}^{\mathrm{BC}}
    \boldsymbol{\mathrm{R}}^{\mathrm{CD}}
    \right]_{00}
    \right|
    } 
    ,
\end{align}
with $\boldsymbol{\zeta}^{\mathrm{BC}} \coloneqq  | \det \boldsymbol{\mathfrak{R}}^{\mathrm{BC}} |^{1/d^{2}}$.
\textbf{b)} Similarly, the quantum obesity \eqref{eq:obesity} of the multi-qubit state in \eqref{eq:p1p2N_R} takes the form
\begin{align}
     \Omega^{p_{1}p_{2\mathrm{N}}}
     =
     \frac{
     \prod_{i=1}^{N}
     \Omega^{p_{2i-1},p_{2i}}
     \zeta^{p_{2i},p_{2i+1}} 
     }{
     \left|
     \left[
    \prod_{i=1}^{N}\mathrm{R}^{p_{2i-1},p_{2i}}\mathfrak{R}^{p_{2i},p_{2i+1}}
    \right]_{00}
    \right|
     }
     ,
\end{align}
with $\zeta^{p_{2i},p_{2i+1}} \coloneqq | \det \mathfrak{R}^{p_{2i},p_{2i+1}} |^{1/4}$ and $\zeta^{p_{2N},p_{2N+1}} \coloneqq 1$.
\label{r:multi_obesity}
\end{corollary}
Overall, these results are telling us that the quantum obesity is a quantity that behaves well under swapping scenarios, following a type of multiplicative law. This means that for given input states and POVM elements, we can readily calculate the amount of quantum obesity which can be extracted by swapping protocols. With these results in place, we now move on to analyse the effect of local filtering operations on the swapping of quantum correlations.

\subsection{
On the role of local filtering operations on the swapping of quantum correlations
}
\label{sec:FS_SF_interplay}

\begin{figure}[h!]
\centering
\includegraphics[width=\columnwidth]{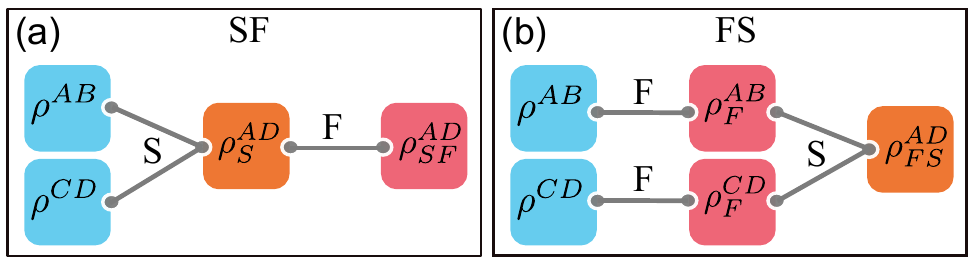}
\vspace{-0.6cm}
\caption{
(Color online) {\bf Two alternatives for an integrated approach to local filtering and swapping}: \textbf{a)} SF approach; initial states $\rho^{\rm AB}$ and $\rho^{\rm CD}$ are first swapped (S) and then local filtering (F) takes place on the post-swapping state. \textbf{a)} FS approach; initial states $\rho^{\rm AB}$ and $\rho^{\rm CD}$ are first independently locally filtered (F) before the swapping (S) protocol takes place.
}
\label{fig:SF_FS}
\end{figure}

In this section we explore the effect of implementing local filtering operations on the swapping of quantum correlations. We propose to evaluate the interplay between swapping and local filtering protocols by considering two alternatives for their application. SF pathway (\autoref{fig:SF_FS}{a}); swapping is implemented on the two initial states $\rho^{AB} \otimes \rho^{CD} \rightarrow \rho^{AD}_S$ and then the post-swapping state is locally filtered as $\rho^{AD}_S \rightarrow \rho^{AD}_{SF}$. FS pathway (\autoref{fig:SF_FS}{b}); the two initial states are first independently filtered as $\rho^{AB} \rightarrow \rho^{AB}_F$ ($\rho^{CD} \rightarrow \rho^{CD}_F$) and then the swapping takes place as $\rho^{AB}_F \otimes \rho^{CD}_F \rightarrow \rho^{AD}_{FS}$. The main goal is to compare the quantum correlations of the final states $ \delta(\rho^{AD}_{SF})$ and $ \delta(\rho^{AD}_{FS})$, for all the correlation measures introduced before $\delta = \text{B}; \text{BF}_3; \text{D}; \text{C}; \Omega$,  \eqref{eq:chsh_measure}-\eqref{eq:obesity}. In the first subsection we prove that these two procedures are equivalent for Bell-diagonal states. Explicitly, that for input Bell-diagonal states and Bell-diagonal POVM element we have $\rho^{AD}_{FS} \equiv \rho^{AD}_{SF}$ and therefore, $\delta(\rho^{AD}_{FS}) \equiv \delta(\rho^{AD}_{SF})$ for all correlation measures.  We then show that this equivalence stops holding true when considering strictly larger families of states. In order to gain some intuition, we first show that $\delta(\rho^{AD}_{FS}) \geq \delta(\rho^{AD}_{SF})$ for partially entangled states with coloured noise (a non-Bell-diagonal state). We then provide numerical evidence for the phenomenon $\delta(\rho^{AD}_{FS}) \geq \delta(\rho^{AD}_{SF})$ holding true for general states in the X-form (a super-class of Bell-diagonal states). Finally, we analytically show that the quantum obesity of almost Bell-diagonal states satisfies $\Omega(\rho^{AD}_{FS}) \geq \Omega(\rho^{AD}_{SF})$. Overall, it is then found that applying filtering \emph{first} is in general more efficient when considering the swapping of quantum correlations.

\subsubsection{Local filtering and swapping: Bell-diagonal states}

We start the exploration of the interplay between the swapping protocol and local filtering by considering input states and measurements to be Bell-diagonal operators. In this regime, we find that the two procedures described above are equivalent as follows.
\begin{result}[Local filtering operations on the swapping of Bell-diagonal states and POVM elements]
    Consider a swapping protocol with input Bell-diagonal states $\rho^{\rm AB}$ and $\rho^{\rm CD}$, and a Bell-diagonal POVM element $E^{\rm BC}$. Then, local filtering operations implemented before or after the swapping takes place lead to the same final state as $\rho^{AD}_{FS} = \rho^{AD}_{SF}$. In particular, this means that the correlations displayed by both procedures are equivalent.
\end{result}

\begin{proof}
    In order to see this equivalence we first check that the swapping protocol takes Bell-diagonal operators into a Bell-diagonal operator. The R-representation of Bell-diagonal states and POVM elements reads:
    {\small \begin{align}
    R^{\rm AB}
    =
    \begin{pmatrix}
    1 & \vec{0}^{T}
    \\
    \vec{0} & T 
    \end{pmatrix}
    ,
    R^{\rm CD}
    =
    \begin{pmatrix}
    1 & \vec{0}^{T}
    \\
    \vec{0} & S
    \end{pmatrix}
    ,
    \mathfrak{R}^{\rm BC}
    =
    \begin{pmatrix}
    A_1 & \vec{0}^{T}
    \\
    \vec{0} & U
    \end{pmatrix}
    .
    \end{align}}
    We can now readily check, using result 1,  that the R-representation of the final state $\rho^{\rm AD}$ then reads:
    \begin{align}
        R^{\rm AD}
        =
        \frac{
        R^{\mathrm{AB}}
        \mathfrak{R}^{\mathrm{BC}}
        R^{\mathrm{CD}}
        }
        {
        \left[
            R^{\mathrm{AB}}
            \mathfrak{R}^{\mathrm{BC}}
            R^{\mathrm{CD}}
        \right]_{00}
        }
        =
        \begin{pmatrix}
        1 & \vec{0}^{T}
        \\
        \vec{0} & TUS
        \end{pmatrix}
        ,
    \end{align}
    meaning that the final state is also Bell-diagonal state. Let us now use this to explore the effect of local filtering operations. On the one hand, considering implementing local filtering first, we have that Bell-diagonal states remain invariant under the action of the KLM-SLOCC and therefore $\rho^{\rm AB}_F = \rho^{\rm AB}$ and $\rho^{\rm CD}_F=\rho^{\rm CD}$, and so after the swapping we get $\rho^{AD}_{FS} = \rho^{AD}_S$. On the other hand, considering swapping first, we have that the swapping protocol maps Bell-diagonal operators into a Bell-diagonal operator. The post-swapping state then remains invariant under the action of KLM-SLOCC as $\rho^{AD}_{SF}=\rho^{AD}_S$, and so effectively achieving $\rho^{AD}_{FS} = \rho^{AD}_{SF}$.
\end{proof}
With this result in place, we then have that quantum correlation measures for both procedures achieve the same values as $\delta(\rho^{AD}_{FS}) \equiv \delta(\rho^{AD}_{SF})$, with $\delta$ being any of the correlation measures considered in this work  \eqref{eq:chsh_measure}-\eqref{eq:obesity}:  $\delta = \text{B}; \text{BF}_3; \text{D}; \text{C}; \Omega$. It is then natural to ask whether one of these two procedures allows for better correlations in general. Since the two procedures are equivalent for Bell-diagonal operators, we now know that we need to move away from this family of states. In order to gain some intuition about this question, let us first address the following non-Bell-diagonal example.

\subsubsection{Local filtering and swapping: non-Bell diagonal example}

\begin{figure}
\centering
\includegraphics[width=\columnwidth]{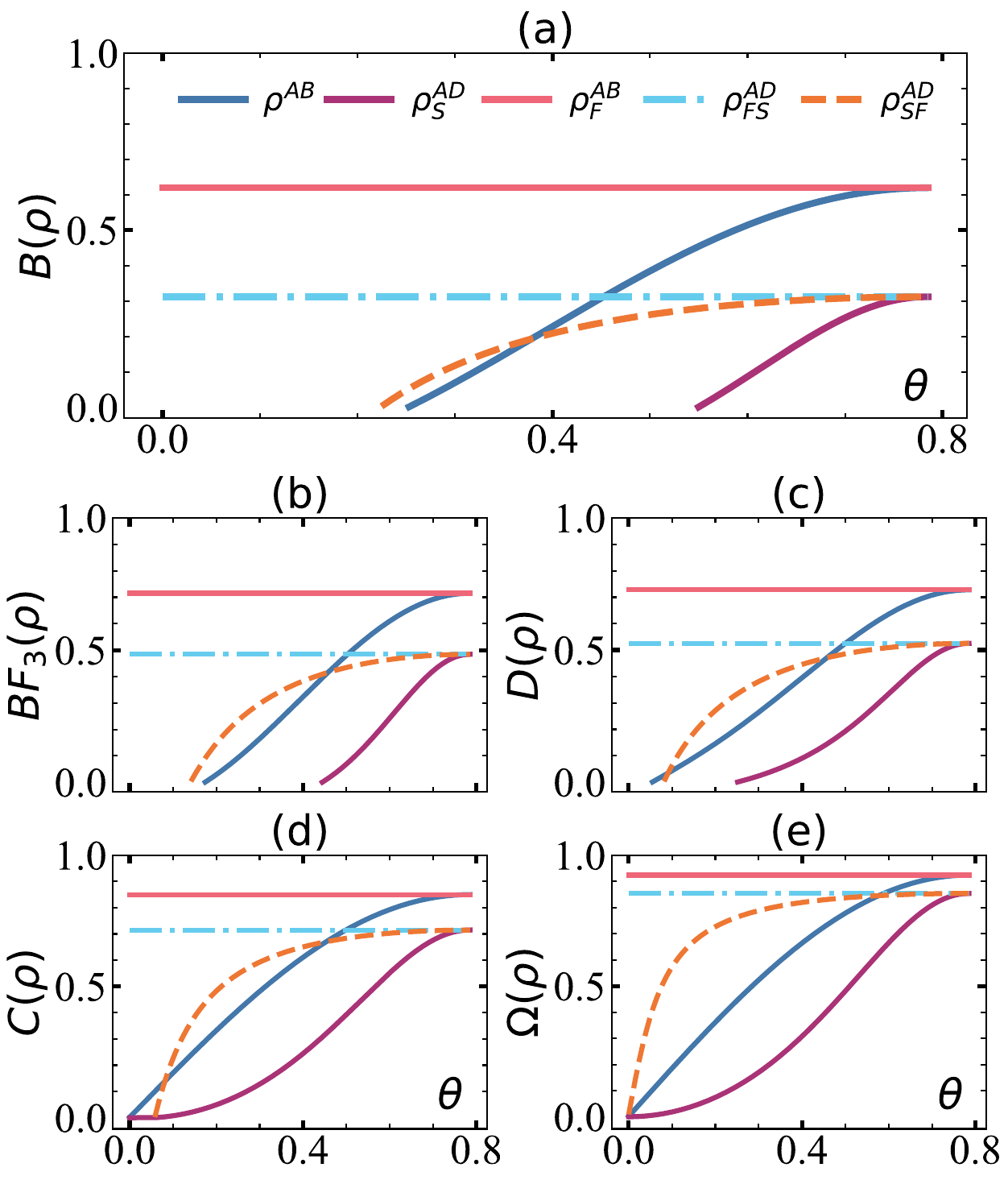}
\caption{(Color online) {\bf SF vs. FS quantum correlations}: (a) Nonlocality, (b) steering, (c) Usefulness for teleportation, (d) concurrence and (e) quantum obesity in terms of $\theta$ for the initial state \eqref{eq:coloured_noise_state} with mixed parameter $p=0.9$. All the considered correlations for $\rho^{\mathrm{AD}}_{\mathrm{FS}}$ outperform  the correlations for $\rho^{\mathrm{AD}}_{\mathrm{SF}}$. Interestingly, the recovered correlations by filtering the post-swapping state surpass the initial correlations for some values of $\theta$.}
\label{fig:case2_state1}
\end{figure}

Consider the following partially entangled state with coloured noise for both initial sources as:
\begin{align}
\rho^{\mathrm{AB}}
=
\rho^{\mathrm{CD}}
=
p
\ketbra{\psi}{\psi}
+
(1-p)
\rho^{\mathrm{A}}
\otimes
(\mathds{1}/2)
,
\label{eq:coloured_noise_state}
\end{align}
with $p\in [0,1]$, $\theta \in [0,\pi/4]$, $\ket{\psi} = \cos{\theta} \ket{00} + \sin{\theta} \ket{11}$, $\rho^{\mathrm{A}} = \tr_{\mathrm{B}}\hspace{2pt}[\ketbra{\psi}{\psi}]$. Consider also the POVM element in BC to be the Bell projector; 
$
\mathrm{E}^{\mathrm{BC}}_2 
= 
\Phi^{\rm BC}_2 
$. In \autoref{fig:case2_state1} we address the quantum correlations of states $\rho^{\mathrm{AB}}$, $\rho^{\mathrm{AB}}_{\mathrm{F}}$, $\rho^{\mathrm{AD}}_{\mathrm{S}}$, $\rho^{\mathrm{AD}}_{\mathrm{SF}}$ and $\rho^{\mathrm{AD}}_{\mathrm{FS}}$ as functions of   $\theta \in [0,\pi/4]$, for $p=0.9$, and analyse the behaviour of these quantum correlation measures as follows.

First, we note that the bare swapping protocol here \emph{degrades} the amount of initial quantum correlations as $\delta (\rho^{AD}_{S}) \leq \delta (\rho^{AB})$, $\delta = \text{B}; \text{BF}_3; \text{D}; \text{C}; \Omega$ (purple curves below blue curves). Second, all swapped quantum correlations are enhanced by implementing local filtering either before or after the swapping takes place as 
$
\delta (\rho^{AD}_{FS}) 
\geq
\delta (\rho^{AB}_S)
$ 
and
$
\delta (\rho^{AD}_{SF}) 
\geq
\delta (\rho^{AB}_S)
$
(cyan and orange curves above purple curve). Third, we note that the filtered state $\rho^{AB}_F$ is overall the most correlated state independently of $\theta$, this, due to the structure of the initial states in \eqref{eq:coloured_noise_state}. Fourth, it is also worth noting that there exist regions ($\theta$) for which $\delta (\rho^{AD}_{SF}) \geq \delta (\rho^{AB})$, for all the correlations here considered. This means that local filtering operations managed to \emph{counteract} the degradation caused by the swapping protocol to the point of (sometimes) even \emph{surpassing} the original amount of correlations $\delta(\rho^{\rm AB})$ (orange curve above blue curve for a region of $\theta$). Fifth and finally, regarding the main phenomenon we wanted to gain intuition about, i. e. the relationship between $\delta (\rho^{AD}_{FS})$ and $ \delta (\rho^{AD}_{SF})$, we have that implementing local filtering \emph{before} the swapping protocol takes place is more efficient than the other way around, specifically, that $\delta (\rho^{AD}_{FS}) \geq \delta (\rho^{AD}_{SF})$ (cyan curves above orange curves). Whilst this is telling us that implementing filters before swapping is indeed more efficient for this particular family of states, one has to be careful with that this is a specific example, and so that in principle it there could exist other families of states displaying different phenomena. With this motivation in mind, the following numerical investigations tell that the phenomenon $\delta (\rho^{AD}_{FS}) \geq \delta (\rho^{AD}_{SF})$ is a generic phenomenon for states in the X-form.

\subsubsection{Local filtering and swapping: X-form states}

\begin{figure}
\centering
\includegraphics[width=\columnwidth]{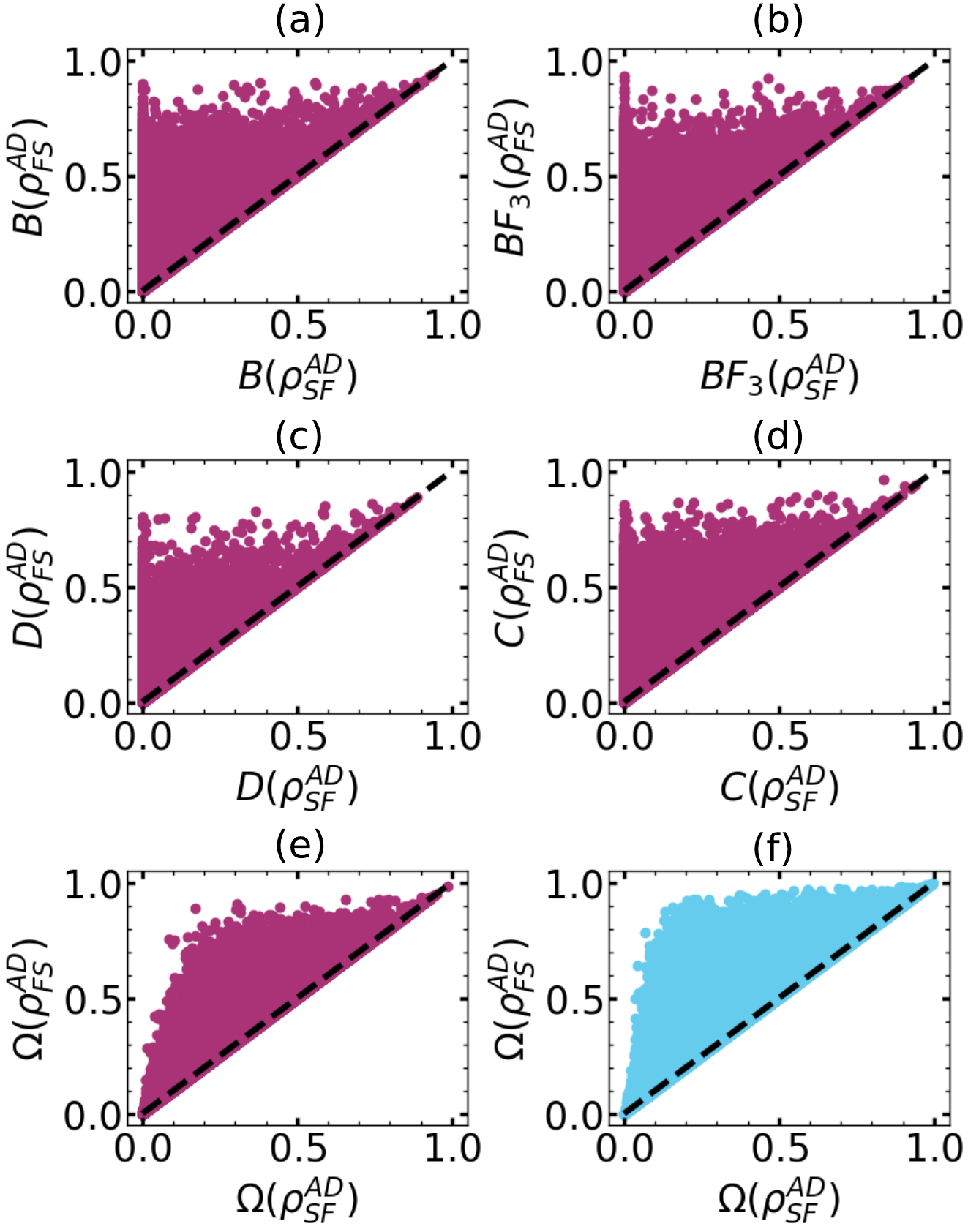}
\vspace{-0.6cm}
\caption{
(Color online) {\bf Quantum correlation measures of random X-form and general two-qubit states}: (a)-(e) $\delta(\rho^{AD}_{FS})$ vs. $\delta(\rho^{AD}_{SF})$, $\delta=\text{B} ,\, \text{BF}_3,\, \text{D},\, \text{C},\, \text{and}\, \Omega$, for the correlation measures of CHSH-nonlocality ($\text{B}(\rho)$), $\rm F3$-steering ($\text{BF}_3(\rho)$), usefulness for teleportation ($\text{D}(\rho)$), entanglement ($\text{C}(\rho)$), and quantum obesity ($\Omega(\rho)$). $10^{6}$ random two-qubit states in the states have been evaluated for each correlation. (f) Quantum obesity for $10^{6}$ random general two-qubit states.
}
\label{fig:correlations_x}
\end{figure}

We now numerically explore the effect of local filtering operations on the swapping of quantum correlation measures within a swapping scenario with $\rho^{\mathrm{AB}} = \rho^{\mathrm{CD}}$, for $10^{6}$ random states in the X-form. In \autoref{fig:correlations_x} the quantum correlation measures of the final states via the two procedures $\delta (\rho^{AD}_{FS})$ vs. $\delta (\rho^{AD}_{SF})$ are plotted: a) CHSH-nonlocality, b) EPR-steering, c) usefulness for teleportation, d) Concurrence and e) quantum obesity. We find that the FS protocol is always greater than the SF protocol, for all the correlation measures here considered. As every correlation was computed independently, it is worth mentioning that the implemented random states are, in principle, different for each correlation. Furthermore, we extend the test of this fact to the set of general states by computing the quantum obesity of both approaches for $10^{6}$ random general states such that $\rho^{\mathrm{AB}} = \rho^{\mathrm{CD}}$ as shown in \autoref{fig:correlations_x} (f). Although we show this general experiment for obesity, similar results (``FS $\geq$ SF'') hold for the rest of correlations (not shown).

Overall, this numerical evidence is then telling us that implementing local filtering \emph{before} the swapping protocol takes place is generally better than the other way around, for all the quantum correlation measures here considered, for all states in the X-form, and so confirming the intuition gained from the partially entangled states with coloured noise explored in the previous subsection \eqref{eq:coloured_noise_state}. Now, one would also like to see whether it is possible to explore this phenomenon in an analytical way. Whilst this is in general a formidable challenge, since it involves several correlation measures as well as general two-qubit states, we give some steps in this regard by considering a specific correlation measure and a specific family of states. In the following section we analytically explore the quantum obesity of the family of almost-Bell-diagonal states.

\subsubsection{Local filtering and swapping: quantum obesity of almost Bell-diagonal states}
\label{sec:FS_SF_ABD_X_states}

We want to explore the relationship between the quantum correlation measures of the final states via the two procedures involving swapping and local filtering $\delta (\rho^{AD}_{FS})$ and $\delta (\rho^{AD}_{SF})$. We have already seen that the two procedures are equivalent for the family of Bell-diagonal states as $\delta (\rho^{AD}_{FS}) = \delta (\rho^{AD}_{SF})$, for all the quantum correlation measures here considered. We have also addressed specific non-Bell diagonal states satisfying $\delta (\rho^{AD}_{FS}) \geq \delta (\rho^{AD}_{SF})$ and, more generally, have provided numerical support for this behaviour to hold true for the family of states in the X-form. Now, in its most general version, one would like to explore this phenomenon for various correlation measures for general two-qubit states. In order to take some further steps in this direction, we now address one specific correlation measure, the quantum obesity, and one family of states, the set of almost-Bell diagonal states. We make this particular choices because we have already seen that the quantum obesity behaves well under swapping, and because the KLM-SLOCC can explicitly be derived for states in the X-form \cite{Matsumura_2020}. Hence, a closed formula for quantum obesity can be derived for states for both SF and FS protocols (i.e.,  $\rho^{\mathrm{AD}}_{\mathrm{SF}}$ and $\rho^{\mathrm{AD}}_{\mathrm{FS}}$ in \autoref{fig:SF_FS}).

Consider a swapping scenario with $\rho^{\mathrm{AB}} = \rho^{\mathrm{CD}}$ arbitrary two-qubit states in the X-form, together with Bell POVM elements on BC. We now want to compare the quantum obesities of the final states after the two protocols $\Omega (\rho^{AD}_{FS})$ and $\Omega (\rho^{AD}_{SF})$. We present the following comparison of the two approaches in terms of the density matrix elements $\rho_{ij}, \; i,j=1,2,3,4$. The quantum obesity of the state $\rho^{\mathrm{AD}}_{\mathrm{FS}}$ can be written as (see \cref{sec:SF_FS}):
\begin{align}
    \Omega^{\mathrm{AD}}_{\mathrm{FS}}
    =
    \Gamma_{\mathrm{FS}}\Omega^{2},
    \label{eq:caso5_result_1}
\end{align}
with $
\Omega
=
\Omega^{\mathrm{AB}}=\Omega^{\mathrm{CD}}
$ 
and $\Gamma_{\mathrm{FS}}$ the following function in terms of the density matrix's elements:
\begin{align}
    \Gamma_{\mathrm{FS}}
    =
    \frac{
    1
    }{
    4
    \left(
    \sqrt{\rho_{11}\rho_{44}}
    +
    \sqrt{\rho_{22}\rho_{33}}
    \right)^{2}
    }.
    \label{eq:caso5_fac1}
\end{align}
Similarly, the quantum obesity of the state $\rho^{\mathrm{AD}}_{\mathrm{SF}}$ reads
\begin{align}
    \Omega^{\mathrm{AD}}_{\mathrm{SF}}
    =
    \Gamma^{(\mathrm{k})}_{\mathrm{SF}}
    \Omega^{2}
    ,
    \label{eq:caso5_result_2}
\end{align}
where $\Gamma^\mathrm{(1)}_{\mathrm{SF}}$ corresponds to a swapping scenario implementing Bell projectors $\{\Phi^{\mathrm{BC}}_{0},\Phi^{\mathrm{BC}}_{1}\}$ whilst $\Gamma^\mathrm{(2)}_{\mathrm{SF}}$ to the corresponds to Bell projectors $\{\Phi^{\mathrm{BC}}_{2},\Phi^{\mathrm{BC}}_{3}\}$. These coefficients are given by
\begin{widetext}
 \begin{align}
&\Gamma^{(1)}_{\mathrm{SF}}=\frac{(\rho^{2}_{11}+\rho_{11}(\rho_{22}+\rho_{33})+\rho_{44}(\rho_{33}+\rho_{44})+\rho_{22}(2\rho_{33}+\rho_{44}))((\rho_{11}+\rho_{44})\sqrt{\rho_{22}\rho_{33}}-\sqrt{(\rho^{2}_{11}+\rho_{22}\rho_{33})(\rho^{2}_{44}+\rho_{22}\rho_{33})})}{2(\rho_{22}\rho_{33}-\rho_{11}\rho_{44})^{2}((\rho_{22}-\rho_{33})^{2}-(\rho_{11}-\rho_{44})^{2}-1)} ,\nonumber\\ 
& \Gamma^{(2)}_{\mathrm{SF}}=\frac{\rho^{2}_{22}+\rho_{22}\rho_{44}+\rho_{33}(\rho_{33}+\rho_{44})+\rho_{11}(\rho_{22}+\rho_{33}+2\rho_{44})}{2(1+(\rho_{22}-\rho_{33})^{2}-(\rho_{11}-\rho_{44})^{2})(\sqrt{\rho_{11}\rho_{44}}(\rho_{22}+\rho_{33})+\sqrt{(\rho^{2}_{22}+\rho_{11}\rho_{44})(\rho^{2}_{33}+\rho_{11}\rho_{44})})} ,
     \label{eq:caso5_fac2}
 \end{align}
\end{widetext}
respectively. Comparing the correlations $\Omega^{\mathrm{AD}}_{\mathrm{FS}}$ and $\Omega^{\mathrm{AD}}_{\mathrm{SF}}$ then reduces to analyse the ratio $\gamma_{k} \coloneqq \Gamma_{\mathrm{FS}}/ \Gamma^{(k)}_{\mathrm{SF}}, \; \Gamma^{(k)}_{\mathrm{SF}} \neq 0$. We have arrived to analytic expressions for these coefficients for arbitrary two-qubit states in the X-form but, it is however not immediately clear if one of the coefficients is larger than the other, so we further consider the subfamily of almost-Bell diagonal states (ABD), and we next explore the behaviour of these ratios for such a family of states.

\begin{figure}[h!]
\centering
\includegraphics[width=\columnwidth]{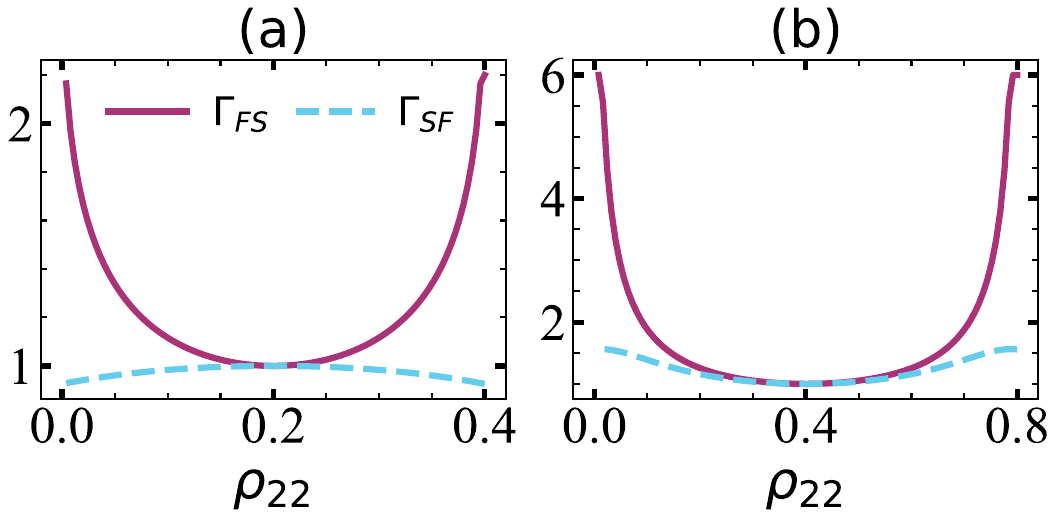}
\vspace{-0.6cm}
\caption{(Color online)  
{\bf Comparing quantum obesity for FS and SF processes}: coefficients $\Gamma_{\mathrm{FS}}$ and $\Gamma_{\mathrm{SF}}$ in terms of $\rho_{22}$ for the functional $\rho_{33}=\alpha-\rho_{22}$ with (a) $\alpha=0.4$ and (b) $\alpha=0.8$. The coefficient $\Gamma_{\mathrm{FS}}$ is always greater or equal than $\Gamma^{(k)}_{\mathrm{SF}}$. 
}
\label{fig:factors}
\end{figure}

As described in the background section, ABD states are X-form states whose R-representation satisfy the additional constraints given by $a_3 = -b_3$ or $a_3 = b_3$. These two constraints in the R-picture translate back to the $\rho$-picture as $\rho_{11} = \rho_{44}$ and $\rho_{22} = \rho_{33}$, respectively. Taking these two subcases into consideration, in \cref{sec:ABD} we show that $\gamma_k \geq 1$ for the set of ABD states. We find that the first subcase ($\rho_{11} = \rho_{44} = \frac{1 -(\rho_{22} + \rho_{33})}{2}$) achieves  $\{\gamma_1=1;\;\gamma_2>1\}$, whilst the second subcase $\rho_{22} = \rho_{33}=\frac{1 -(\rho_{11} + \rho_{44})}{2}$ in turn achieves  $\{\gamma_2=1;\;\gamma_1>1\}$, and so both cases effectively achieving $\gamma_k\geq1$, $k = 1,2$. In \autoref{fig:factors}(a)-(b) we present an example of the behavior of $\gamma_k$ for the case $\rho_{11}=\rho_{44}$ ($a_3=-b_3$) and the functional $\rho_{33}=\alpha-\rho_{22}$, from which we get that $\Gamma_{\rm FS} \geq \Gamma_{\rm SF}$. The analysis in terms of the two parameters $\rho_{22}$ and $\rho_{33}$ can be followed in \cref{sec:ABD}.

\section{Conclusions} 
\label{sec:conclusions}

In this work we explored the swapping of general quantum correlations as well as the effect of local filtering operations, or stochastic local operations and classical communication (SLOCC), on the swapping of such quantum correlations. We considered correlation measures for the quantum properties of: Bell-nonlocality, EPR-steering, usefulness for teleportation, entanglement and quantum obesity.  

In the first part of this work we addressed the swapping protocol without local filtering operations. In this regime we showed how the quantum correlations of arbitrary input quantum states can \emph{fully} be preserved when the general input state is swapped together with arbitrary combinations of Bell states and Bell measurements. The analytical construction we present here explains the \emph{numerical} data found in \cite{s8} for general two-qubit states, and extends the analytical arguments also shown in \cite{s8} for the particular case of the \emph{concurrence} of input states in the \emph{X-form}. Specifically, our analytical construction holds true for \emph{general} two-qubit states (X-form states included) as well as for \emph{general} quantum correlation measures which are invariant under local unitary operations and so including all the measures addressed in this work (concurrence included). Furthermore, we derived an explicit formula for the quantum obesity of the final post-swapping state in terms of the quantum obesities of general input states, and therefore establishing an upper limit at which quantum obesity can effectively be swapped. The formula we derived in this regime can naturally be extended to more elaborate swapping scenarios involving higher dimensional systems like sources emitting two-qudit systems as well as multi-source chain of $N$-qubit systems.

In the second part of this work we explored the effect of local filtering operations on the swapping of quantum correlations. In this regard, we addressed two scenarios, one in which the local filtering operation is independently implemented on both input states $\rho^{\rm AB}$ and $\rho^{\rm CD}$ \emph{before} the swapping protocol takes place (final state denoted as $\rho^{AD}_{FS}$), and the other way around, a scenario where the local filtering takes place \emph{after} the swapping is performed (final state denoted as $\rho^{AD}_{SF}$), and explored the relationship between the quantum correlations these two procedures achieve $\delta (\rho^{AD}_{SF})$ and $\delta (\rho^{AD}_{SF})$, with $\delta$ representing correlation measures for the quantum properties of interest. We first show that these two scenarios are completely equivalent for the family of Bell-diagonal states, for all quantum correlations here considered. We then show that this equivalence stops holding true when considering the strictly larger family of almost Bell-diagonal states. In this regard, we provided explicit examples of states (partially entangled states with coloured noise) for which implementing local filtering operations \emph{before} swapping is more efficient than the other way around $\delta (\rho^{AD}_{FS}) \geq \delta (\rho^{AD}_{SF})$. These states in particular allow for the local filtering operations to \emph{counteract} the degradation caused by the swapping protocol, sometimes even achieving larger values than the initial amount of correlations as $\delta (\rho^{AD}_{SF}) \geq \delta (\rho^{AB})$. Moreover, we provided numerical evidence supporting the phenomenon of implementing local filtering operations first being more efficient $\delta (\rho^{AD}_{FS}) \geq \delta (\rho^{AD}_{SF})$, for general two-qubit states in the X-form, and for all the quantum correlations under consideration. Finally, we analytically showed that this holds true for the quantum obesity of almost-Bell-diagonal states. 

\section*{Acknowledgements}
\label{sec:acknowledgements}
A.F.D. acknowledges financial support from the International Research Unit of Quantum Information, Kyoto University, the Center for Gravitational Physics and Quantum Information (CGPQI), COLCIENCIAS (Grant 756-2016), and UK EPSRC (EP/L015730/1). P.R and C.E.S. acknowledge funding from Universidad de C\'ordoba (Grants FCB-08-19 and FCB-09-22). P.R is also thankful for the support from CNPq Grant No. 130267/2022-8 and FAPESP Grant No.   	
2022/12382-4.

\appendix
\onecolumngrid
\section{Proof of \cref{r:general_swapping}}
\label{app:proof_general_swapping}

\begin{proof}
The unnormalised post-swapping state $\widetilde{\rho}^{AD}$ can be written as follows:
\begin{align}
 \widetilde{\rho}^{\mathrm{AD}}
 &
 = \tr_{\mathrm{BC}}
 \left[
     \left(
        \rho^{\rm AB}
        \otimes
        \rho^{\rm CD}
     \right)
     \left(
        \mathds{1}^{\rm A}
         \otimes
         E^{\rm BC}
         \otimes
         \mathds{1}^{\rm D}
     \right)
 \right]
 ,
 \nonumber\\
 &
 {=}
 \tr_{\mathrm{BC}}\hspace{2pt}
 \Biggl[
 \left(
 \frac{1}{4}\sum_{i,j}\mathrm{R}^{\mathrm{AB}}_{ij}\sigma^{\mathrm{A}}_{i}\otimes \sigma^{\mathrm{B}}_{j} \otimes \frac{1}{4}\sum_{m,n}\mathrm{R}^{
\mathrm{CD}}_{mn}\sigma^{\mathrm{C}}_{m}\otimes \sigma^{\mathrm{D}}_{n}
\right)
\left(
\mathbb{1}^{\mathrm{A}}
\otimes
\sum_{k,l}\mathfrak{R}^{\mathrm{BC}}_{kl}\sigma^{\mathrm{B}}_{k}\otimes \sigma^{\mathrm{C}}_{l} \otimes \mathbb{1}^{\mathrm{D}}
\right)\Biggr],
\nonumber\\
&
{=}
\tr_{\mathrm{BC}}
\Biggl[
\frac{1}{16}
\sum_{i,j}\sum_{m,n}\sum_{k,l}\mathrm{R}^{\mathrm{AB}}_{ij}\mathrm{R}^{
\mathrm{CD}}_{mn}\mathfrak{R}^{\mathrm{BC}}_{kl}
\sigma^{\mathrm{A}}_{i}\otimes \sigma^{\mathrm{B}}_{j}\sigma^{\mathrm{B}}_{k} \otimes \sigma^{\mathrm{C}}_{m}\sigma^{\mathrm{C}}_{l}\otimes \sigma^{\mathrm{D}}_{n}
\Biggr],
\nonumber\\
&
{=}
\frac{1}{4}\sum_{i,j}\sum_{m,n}\sum_{k,l}\delta_{jk}\delta_
 {ml}\mathrm{R}^{\mathrm{AB}}_{ij}\mathrm{R}^{
\mathrm{CD}}_{mn}\mathfrak{R}^{\mathrm{BC}}_{kl}\sigma^{\mathrm{A}}_{i}\otimes \sigma^{\mathrm{D}}_{n},
 \;\; \text{using}\; \tr\left[\sigma_{i}\sigma_{j}\right]=2\delta_{ij}.
 \nonumber\\ 
&
{=}
\frac{1}{4}
\sum_{i,n}\sum_{j,m}\mathrm{R}^{\mathrm{AB}}_{ij}\mathrm{R}^{\mathrm{CD}}_{mn}\mathfrak{R}^{\mathrm{BC}}_{jm}\sigma^{\mathrm{A}}_{i}\otimes \sigma^{\mathrm{D}}_{n},
\nonumber\\
&
{=}
\frac{1}{4}
\sum_{i,n}
(\mathrm{R}^{\mathrm{AB}}\mathfrak{R}^{\mathrm{BC}
}\mathrm{R}^{\mathrm{CD}})_{in}\ \sigma^{\mathrm{A}}_{i}\otimes \sigma^{\mathrm{D}}_{n},
\nonumber\\
&
{=}
\frac{
1
}{
4
}
\sum_{i,n=0}^{3}
\widetilde{\mathrm{R}}^{\mathrm{AD}}_{in}
\sigma^{\mathrm{A}}_{i}
\otimes \sigma^{\mathrm{D}}_{n}, \;\; 
\text{defining}\; 
\widetilde{\mathrm{R}}^{\mathrm{AD}}
\coloneqq
\mathrm{R}^{\mathrm{AB}}\mathfrak{R}^{\mathrm{BC}}\mathrm{R}^{\mathrm{CD}}
.
\label{eq:proof_1}
\end{align}
The proof ends with the state normalisation; $\tr [ \widetilde{\rho}^{AD} ] = [\widetilde{\mathrm{R}}^{\mathrm{AD}}]_{00}$, $[X]_{00}$ is the (0,0) component of matrix $X$. The normalised post-swapping state then reads $\rho^{\mathrm{AD}}=\widetilde{\rho}^{\mathrm{AD}}/[\widetilde{\mathrm{R}}^{\mathrm{AD}}]_{00}$.
\end{proof}

\section{Details about \cref{cr:standard_swapping}}
\label{app:proof_standard_swapping}
\begin{proof}
For a general state in part AB, considering any Bell state in part CD and any Bell projector as a POVM element in part BC 
\begin{align}
    \rho^{\mathrm{CD}}
    =
    \Phi^{\mathrm{CD}}_n
    =
     (\mathbb{1}\otimes \sigma_{n})\left(\frac{1}{4}\sum_{k,l}\mathrm{R}^{
\mathrm{CD}}_{kl}\sigma^{\mathrm{C}}_{k}\otimes \sigma^{\mathrm{D}}_{l}\right)(\mathbb{1}\otimes \sigma_{n})
\end{align}
and 
\begin{align}
    E^{\mathrm{BC}}
    =
    \Phi^{\mathrm{BC}}_m
    =
     (\mathbb{1}\otimes \sigma_{m})
     \left(
     \sum_{r,s}\mathrm{R}^{
\mathrm{BC}}_{rs}\sigma^{\mathrm{B}}_{r}\otimes \sigma^{\mathrm{C}}_{s}\right)(\mathbb{1}\otimes \sigma_{m}) ,
\end{align}
respectively. The resulting state after swapping reads:
\begin{align}
    \widetilde{\rho}^{\mathrm{AD}}
 &{=}
 \tr_{\mathrm{BC}}
 \left[
     \left(
        \rho^{\rm AB}
        \otimes
        \Phi^{\mathrm{CD}}_n
     \right)
     \left(
        \mathds{1}^{\rm A}
         \otimes
         \Phi^{\mathrm{BC}}_m
         \otimes
         \mathds{1}^{\rm D}
     \right)
 \right]
 ,\\
&{=}
\tr_{\mathrm{BC}}
\Biggl[
\frac{1}{64}\sum_{i,j,k,l,r,s}\mathrm{R}^{\mathrm{AB}}_{ij}\mathrm{R}^{
\mathrm{CD}}_{kl}\mathrm{R}^{\mathrm{BC}}_{rs}
\sigma^{\mathrm{A}}_{i}\otimes \sigma^{\mathrm{B}}_{j}\sigma^{\mathrm{B}}_{r} \otimes \sigma^{\mathrm{C}}_{k}\sigma_{m}\sigma^{\mathrm{C}}_{s}\sigma_{m}\otimes \sigma_{n}\sigma^{\mathrm{D}}_{l}\sigma_{n}
\Biggr],\\
&{=}
\frac{1}{64}\sum_{i,j,k,l,r,s}\mathrm{R}^{\mathrm{AB}}_{ij}\mathrm{R}^{
\mathrm{CD}}_{kl}\mathrm{R}^{\mathrm{BC}}_{rs}
\sigma^{\mathrm{A}}_{i}\otimes \tr
\left[\sigma^{\mathrm{B}}_{j}\sigma^{\mathrm{B}}_{r}\right] \otimes \tr
\left[\sigma^{\mathrm{C}}_{k}\sigma_{m}\sigma^{\mathrm{C}}_{s}\sigma_{m}\right]\otimes \sigma_{n}\sigma^{\mathrm{D}}_{l}\sigma_{n},
\label{eq:corollary_trace_BC}
\end{align}
Remembering that $\mathrm{R}^{\mathrm{BC}}$ and $\mathrm{R}^{\mathrm{CD}}$ correspond to Bell states and taking into account the following properties:
\begin{align}
    &\tr\hspace{2pt}
    \left[
    \sigma_{j}\sigma_{r}
    \right]
    =
    2\delta_{jr}
    ,
    \hspace{0.5cm}
    j,r=0,1,2,3,
    \hspace{0.5cm}
    \sigma_{0}=\mathbb{1}_{2}
    .\\
    & \tr\hspace{2pt}
    \left[
    \sigma_{k}\sigma_{m}\sigma_{s}\sigma_{m}
    \right]
    =2(\delta_{km}\delta_{sm}-\delta_{ks}+\delta_{km}\delta_{ms})+4(\delta_{ks}\delta_{0m}\delta_{0m}+\delta_{0k}\delta_{0s})-8\delta_{0k}\delta_{0s}\delta_{0m}\delta_{0m}+\sum_{k,m,s}\mathcal{E}_{0kms}\delta_{0m}
    ,
    \label{eq:trpauli4}
\end{align}
where $\mathcal{E}$ is the Levi-Civita symbol. Lastly \eqref{eq:corollary_trace_BC} leads to:
\begin{align}
    \widetilde{\rho}^{\mathrm{AD}}
 &{=}
(\mathbb{1}\otimes \sigma_{n})
\left(\frac{1}{16}\sum_{i,j}\mathrm{R}^{\mathrm{AB}}_{ij}\sigma^{\mathrm{A}}_{i}\otimes \sigma_{m}\sigma^{\mathrm{D}}_{j}\sigma_{m}\right)(\mathbb{1}\otimes \sigma_{n}),\\
 &{=}
(\mathbb{1}\otimes \sigma_{n}\sigma_{m})
\left(\frac{1}{16}\sum_{i,j}\mathrm{R}^{\mathrm{AB}}_{ij}\sigma^{\mathrm{A}}_{i}\otimes \sigma^{\mathrm{D}}_{j}\right)(\mathbb{1}\otimes \sigma_{m}\sigma_{n}),
\label{eq:corollary_normalise}
\end{align}
The statement \cref{cr:standard_swapping} takes place by normalising the state \eqref{eq:corollary_normalise}. 
\end{proof}

\section{Proof of \cref{r:qudit_swapping}}
\label{app:proof_qudit_swapping}
\begin{proof}
\cref{r:general_swapping} can be extended to higher dimensional Hilbert spaces $D(\mathds{C}^d \otimes \mathds{C}^d)$. For doing so, let us consider the following two initial two-qudit states
\begin{align}
    \boldsymbol{\rho}^{AB}
    =
    \frac{1}{d^{2}}
    \sum_{i,j=0}^{d^{2}-1}
    \mathbf{R}^{AB}_{ij}
    \boldsymbol{\sigma}^{A}_{i}
    \otimes 
    \boldsymbol{\sigma}^{B}_{j} 
    ,\hspace{0.7cm}
    \boldsymbol{\rho}^{CD}
    =
    \frac{1}{d^{2}}
    \sum_{m,n=0}^{d^{2}-1}
    \mathbf{R}^{CD}_{mn}
    \boldsymbol{\sigma}^{C}_{m}
    \otimes 
    \boldsymbol{\sigma}^{D}_{n} 
    .
\end{align}
where $\boldsymbol{\sigma}_{0}=\mathbb{1}_{d}$ and $\{\boldsymbol{\sigma_{i}}\}_{i=1}^{d^{2}-1}$ is a set of ($d\times d$)-operators satisfying {\cite{Jing_2022_1,Br_ning_2012}}
\begin{align}
\tr
\left[\boldsymbol{\sigma}_{i}\boldsymbol{\sigma}_{j}\right]=d\delta_{ij},\ \ \ \tr\hspace{2pt}
\left[
\boldsymbol{\sigma}_{i}
\right]=0,\ \ \ \ i,j=0,1,...,d^{2}-1 .
 \end{align}
Let us also consider the following POVM element to carry out the measurement on part BC:
\begin{align}
  \boldsymbol{E}^{BC}= \sum_{k,l=0}^{d^{2}-1}\boldsymbol{\mathfrak{R}}^{\mathrm{BC}}_{kl}\boldsymbol{\sigma}^{B}_{k}\otimes \boldsymbol{\sigma}^{C}_{l} .
\end{align}
With the associate matrices  $\mathbf{R}^{AB}_{ij}=\tr\hspace{2pt}\left[\boldsymbol{\rho}^{AB}(\boldsymbol{\sigma}_{i} \otimes \boldsymbol{\sigma}_{j} )\right]$, $\mathbf{R}^{CD}_{mn}=\tr\hspace{2pt}\left[\boldsymbol{\rho}^{CD}(\boldsymbol{\sigma}_{m} \otimes \boldsymbol{\sigma}_{n} )\right]$ and $\boldsymbol{\mathfrak{R}}^{\mathrm{BC}}_{kl}=\tr\hspace{2pt}\left[\boldsymbol{E}^{BC}(\boldsymbol{\sigma}_{k} \otimes \boldsymbol{\sigma}_{l} )\right]/d^{2}$, the proof is completed by computing the post-swapping state $\boldsymbol{\rho}^{AD}$ in a similar way than that in \cref{app:proof_general_swapping}.
\end{proof}

\section{Proof of  \cref{r:multiq_swapping}}
\label{sec:chainqubits}

\begin{figure}[h!]
\centering
\includegraphics[scale=0.5]{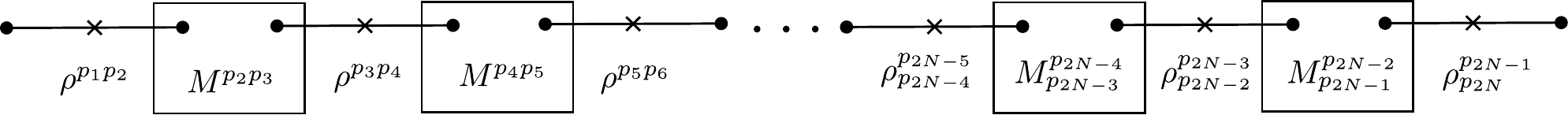}
\caption{{\bf Multi-qubit swapping}: $\mathrm{N}$ sources with two qubits are considered.}
\label{fig:swapping_chain}
\end{figure}

\begin{proof}
Let the input state $\rho^{p_1...p_{2N}}$ and the measurement $M^{p_2...p_{2N-1}}$ be as in  \eqref{eq:multiq_measure_chain}. 
\begin{align}
    \rho^{p_1...p_{2N}}
    \coloneqq
    \bigotimes_{i=1}^{N}
    \rho^{p_{2i-1},p_{2i}}
    , \hspace{0.6cm}
    M^{p_2...p_{2N-1}}
    \coloneqq
    \bigotimes_{j=1}^{N-1}
     M^{p_{2j},p_{2j+1}}
     ,
\end{align}
with the bipartite states and operators as described in \autoref{fig:swapping_chain}. The state (up to normalisation) after the swapping protocol is given by:
\begin{align}
 &
 \widetilde{\rho}^{\,p_{1},p_{2N}}
 \nonumber
 \\
 &
 {=}
 \tr_{
 \vec p 
 / 
 (p_{1}, p_{2N})
 }
 \hspace{-0.1cm}
 \left[
     \left(
        \bigotimes_{i=1}^{N}
        \rho^{p_{2i-1},p_{2i}}
     \right)
     \hspace{-0.1cm}
     \left(
        \mathds{1}^{p_{1}}
        \otimes
        \left(
         \bigotimes_{j=1}^{N-1}
         M^{p_{2j},p_{2j+1}}
         \right)
         \otimes
         \mathds{1}^{p_{2N}}
     \right)
 \right]
 ,
 \\
 &
 {=}
 \frac{1}{4^{N}}
 \tr_{
 \vec p 
 / 
 (p_{1}, p_{2N})
 }
 \hspace{-0.1cm}
 \left[
     \left(
        \bigotimes_{i=1}^{N}
        \sum_{k_i,l_i}
        R_{k_il_i}^{p_{2i-1},p_{2i}}
        \,
        \sigma^{p_{2i-1}}_{k_i}
        \otimes
        \sigma^{p_{2i}}_{l_i}
     \right)
     \hspace{-0.1cm}
     \left(
        \mathds{1}^{p_{1}}
        \otimes
        \left(
             \bigotimes_{j=1}^{N-1}
             \sum_{m_j,n_j}
             M_{m_jn_j}^{p_{2j},p_{2j+1}}
             \sigma^{p_{2j}}_{m_j}
             \otimes
             \sigma^{p_{2j+1}}_{n_j}
         \right)
         \otimes
         \mathds{1}^{p_{2N}}
     \right)
 \right]
 ,
 \\
 &
 {=}
 \frac{1}{4^{N}}
 \tr_{
 \vec p 
 / 
 (p_{1}, p_{2N})
 }
 \hspace{-0.1cm}
 \Bigg[
    \sum_{k_i,l_i}
    R_{kl}
        \,
     \left(
        \bigotimes_{i=1}^{N}
        \sigma^{p_{2i-1}}_{k_i}
        \otimes
        \sigma^{p_{2i}}_{l_i}
     \right)
     \left(
        \mathds{1}^{p_{1}}
        \otimes
        \left(
            \sum_{m_j,n_j}
            M_{mn}
             \bigotimes_{j=1}^{N-1}
             \sigma^{p_{2j}}_{m_j}
             \otimes
             \sigma^{p_{2j+1}}_{n_j}
         \right)
         \otimes
         \mathds{1}^{p_{2N}}
     \right)
 \Bigg]
 ,
\end{align}
where we have defined $R_{kl}
    \coloneqq
    \prod_{i=1}^{N}
        R_{k_il_i}^{p_{2i-1},p_{2i}}$ and $M_{mn}
    \coloneqq
    \prod_{j=1}^{N-1}
         M_{m_jn_j}^{p_{2j},p_{2j+1}}$. Taking into account that
\begin{align}
     \left(
        \bigotimes_{i=1}^{N}
        \sigma^{p_{2i-1}}_{k_i}
        \otimes
        \sigma^{p_{2i}}_{l_i}
     \right)
     &=
     \left(
        \sigma^{p1}_{k_1}
        \otimes
         \sigma^{p_2}_{l_1}
        \otimes
        \left[
            \bigotimes_{i=2}^{N-1}
            \,
            \sigma^{p_{2i-1}}_{k_i}
            \otimes
            \sigma^{p_{2i}}_{l_{i}}
        \right]
        \otimes
        \sigma^{p_{2N-1}}_{k_{N}}
        \otimes
        \sigma^{p_{2N}}_{l_{N}}
     \right)
 ,\nonumber
    \\
    &=
    \sigma^{p1}_{k_1}
    \otimes
    \left(
        \bigotimes_{i=1}^{N-1}
        \sigma^{p_{2i}}_{l_{i}}
        \otimes
        \sigma^{p_{2i+1}}_{k_{i+1}}
    \right)
    \otimes
    \sigma^{p_{2N}}_{l_{N}}
    ,\nonumber
\end{align}
and the trace identity
\begin{align}
    \tr
    \left[
         \bigotimes_{i=1}^{N-1}
         \sigma^{p_{2i}}_{l_i}
         \sigma^{p_{2j}}_{m_i}
         \otimes
         \sigma^{p_{2i+1}}_{k_{i+1}}
         \sigma^{p_{2j+1}}_{n_i}
     \right]
     =
     \prod_{i=1}^{N-1}
     4\,
     \delta_{li,m_i}
     \,
     \delta_{k_{i+1},n_{i}}
     =
     4^{N-1}
     \prod_{i=1}^{N-1}
     \delta_{li,m_i}
     \,
     \delta_{k_{i+1},n_{i}}
     .\nonumber
\end{align}
The post-swapping state finally reads:
\begin{align}
 &
 \widetilde{\rho}^{\,p_{1},p_{2N}}
 \nonumber\\
 &
 {=}
 \frac{1}{4^{N}}
 \tr_{
 \vec p 
 / 
 (p_{1}, p_{2N})
 }
 \hspace{-0.1cm}
 \left[
    \sum_{k_i,l_i}
    \sum_{m_i,n_i}
    R_{kl}
    M_{mn}
     \left(
        \sigma^{p_{1}}_{k_{1}}
        \otimes
        \left(
             \bigotimes_{i=1}^{N-1}
             \sigma^{p_{2i}}_{l_i}
             \sigma^{p_{2j}}_{m_i}
             \otimes
             \sigma^{p_{2i+1}}_{k_{i+1}}
             \sigma^{p_{2j+1}}_{n_i}
         \right)
         \otimes
         \sigma^{p_{2N}}_{l_{2N}}
     \right)
 \right]
 ,\nonumber\\
 &
 {=}
 \frac{1}{4}
    \sum_{k_i,l_i}
    \sum_{m_i,n_i}
    R_{kl}
    M_{mn}
    \left(
    \prod_{i=1}^{N-1}
     \delta_{li,m_i}
     \,
     \delta_{k_{i+1},n_{i}}
     \right)
        \sigma^{p_{1}}_{k_{1}}
         \otimes
         \sigma^{p_{2N}}_{l_{2N}}
 ,
 \nonumber\\
 &
 {=}
 \frac{1}{4}
\sum_{k_{1},l_{2N}}
\left(
    \sum_{\vec k \vec l/(k_1l_{2N})}
    R_{k,l}
    M_{l,k+1}
 \right)
    \sigma^{p_{1}}_{k_{1}}
     \otimes
     \sigma^{p_{2N}}_{l_{2N}}
 ,
\nonumber\\
 &
 {=}
 \frac{1}{4}
\sum_{k_{1},l_{2N}}
\left(
    \sum_{\vec k \vec l/(k_1l_{2N})}
    \left(
    \prod_{i=1}^{N}
         R_{k_il_i}^{p_{2i-1},p_{2i}}
     \right)
     \left(
     \prod_{i=1}^{N-1}
         M_{l_ik_{i+1}}^{p_{2i},p_{2i+1}}
     \right)
 \right)
    \sigma^{p_{1}}_{k_{1}}
     \otimes
     \sigma^{p_{2N}}_{l_{2N}}
 ,
 \nonumber\\
 &
 {=}
 \frac{1}{4}
\sum_{k_{1},l_{2N}}
\left(
    \sum_{\vec k \vec l/(k_1l_{2N})}
    \left(
    \prod_{i=1}^{N-1}
         R_{k_il_i}^{p_{2i-1},p_{2i}}
         M_{l_ik_{i+1}}^{p_{2i},p_{2i+1}}
    \right)
         R_{k_Nl_N}^{p_{2N-1},p_{2N}}
 \right)
    \sigma^{p_{1}}_{k_{1}}
     \otimes
     \sigma^{p_{2N}}_{l_{2N}}
 ,
 \nonumber\\
&
{=}
\frac{1}{4}
\sum_{k_{1},l_{2N}}
\left(
\prod_{i=1}^{N}
    \mathrm{R}^{p_{2i-1},p_{2i}}
    \,
    \mathfrak{R}^{p_{2i},p_{2i+1}}
\right)_{k_{1}l_{2N}}
\sigma^{p_1}_{k_1}
\otimes 
\sigma^{p_{2N}}_{l_{2N}}
,
\nonumber\\
&
{=}
\frac{1}{4}
\sum_{k_{1},l_{2N}}
\widetilde{\mathrm{R}}^{p_{1},p_{2N}}_{k_{1}l_{2N}}
\,
\sigma^{p_1}_{k_1}
\otimes 
\sigma^{p_{2N}}_{l_{2N}}
,
\end{align}
where we have defined the operator:
\begin{align}
    \widetilde{\mathrm{R}}^{p_{1},p_{2N}}
    \coloneqq
    \prod_{i=1}^{N}
    \mathrm{R}^{p_{2i-1},p_{2i}}
    \,
    \mathfrak{R}^{p_{2i},p_{2i+1}}
    ,
    \;\;
    \text{with}
    \;\;
    \mathfrak{R}^{p_{2N},p_{2N+1}}
    \coloneqq
    \mathds{1}^{p_{2N},p_{2N+1}}
    .
    \label{eq:multiq_demo}
\end{align}
After normalising \eqref{eq:multiq_demo} we get the statement in \eqref{eq:p1p2N_R}.
\end{proof}

\section{Details on Corollary 4}

\begin{proof} 
Using \eqref{eq:r1}, the quantum obesity \eqref{eq:obesity} takes the form:
\begin{align}
    \nonumber\Omega^{\mathrm{AD}}&
    =
    \left|
    \det
    R^{\rm AD}
    \right|^{1/4}
    =
    \left|
    \det
    \frac{
    R^{\rm AB}
    \mathfrak{R}^{\rm BC}
    R^{\rm CD}
    }{
    \left[ 
    R^{\rm AB}
    \mathfrak{R}^{\rm BC}
    R^{\rm CD}
    \right]_{00}
    }
    \right|^{1/4}
    =
    \frac{
    \left|
    \det
    R^{\rm AB}
    \mathfrak{R}^{\rm BC}
    R^{\rm CD}
    \right|^{1/4}
    }{
    \left[ 
    R^{\rm AB}
    \mathfrak{R}^{\rm BC}
    R^{\rm CD}
    \right]_{00}
    }
    =
    \frac{
    \Omega^{\mathrm{AB}}
    \zeta^{\mathrm{BC}}
    \Omega^{\mathrm{CD}}
    }{
    \left[ 
    R^{\rm AB}
    \mathfrak{R}^{\rm BC}
    R^{\rm CD}
    \right]_{00}
    }
    . 
    \label{eq:caso4_obesity1}
\end{align}
\end{proof}

\section{Quantum obesity throughout SF and FS pathways for the set of states in the X-form}\label{sec:SF_FS}

\subsection{SF pathway}\label{sec:appendix_5_A}
The {\it locally-filtered-post-swapping} state in the R-representation is given:
\begin{align}
\mathrm{R}^{\mathrm{AD}}_{\mathrm{SF}}=\frac{\Lambda_{A}\mathrm{R}^{\mathrm{AD}}\Lambda^{T}_{B}\det{f_{A}}\det{f_{B}}}{\tr\hspace{2pt}\left[(f^{\dagger}_{A}f_{A}\otimes f^{\dagger}_{B}f_{B})\rho^{\mathrm{AD}}\right]},
\end{align}
where $\mathrm{R}^{\mathrm{AD}}$ is the post-swapping state from \cref{r:general_swapping}, $f_{\mathrm{A}}$, $f_{\mathrm{B}}$ are the optimal filters for $\rho^{AD}$ satisfying
 $f_{i}f_{i}\leq \mathbb{1}$ and the operators $\Lambda_{i}=\Upsilon (f_{i}\otimes f^{*}_{i})\Upsilon^{\dagger}/|\det{f_{i}}|$ are the orthochronous Lorentz transformations with
 $\det{\Lambda_{i}}=1\ \ \ \text{and} \ \  \Lambda^{00}_{i}\geq0$ {\cite{Verstraete_2001_2,pal_2015}}. Similarly, $\Upsilon$ is given by
 \begin{align}
\Upsilon=\frac{1}{\sqrt{2}}\begin{pmatrix}
1 & 0 & 0&1\\
0 &1&1&0\\
0 &i & -i &0\\
1& 0 & 0& -1
\end{pmatrix}.
 \end{align}
Therefore, the quantum obesity for the state after the SF pathway  is given by
\begin{align}
 \nonumber \Omega^{\mathrm{AD}}_{\mathrm{SF}}&=\left|\det{\frac{\Lambda_{A}\mathrm{R}^{\mathrm{AD}}\Lambda^{T}_{B}\det{f_{\mathrm{A}}}\det{f_{\mathrm{B}}}}{\tr\hspace{2pt}\left[(f^{\dagger}_{\mathrm{A}}f_{\mathrm{B}}\otimes f^{\dagger}_{\mathrm{B}}f_{\mathrm{B}})\rho^{\mathrm{AD}}\right]}}\right|^{1/4},\\ 
&=\frac{\left|\det{f_{\mathrm{A}}}\det{f_{\mathrm{B}}}\right|}{\left|4\widetilde{\mathrm{R}}^{AD}_{(00)}\tr\hspace{2pt}\left[(f^{\dagger}_{\mathrm{A}}f_{\mathrm{A}}\otimes f^{\dagger}_{\mathrm{B}}f_{\mathrm{B}})\rho^{\mathrm{AD}}\right]\right|}\Omega^{\mathrm{AB}}\Omega^{\mathrm{BC}}\Omega^{\mathrm{CD}}.
    \label{eq:caso5_OB1}
\end{align}

To get a close formula for the quantum obesity after this SF pathway, we consider the following: states in the set XFS for both sources, for which local filters are explicitly known {\cite{Matsumura_2020}}. POVM  belonging to the Bell-basis such that $\Omega^{\mathrm{BC}}=1$. Under these assumptions, $\Omega^{\mathrm{AB}}=\Omega^{\mathrm{CD}}=\Omega$, and $\Omega^{\mathrm{AD}}_{\mathrm{SF}}$ reads
\begin{align}
 &\Omega^{\mathrm{AD}}_{\mathrm{SF}}
 =
 \Gamma^{(k)}_{\mathrm{SF}}
 \Omega^{2},
\label{eq:caso5_OB2}
\end{align}
where $\Gamma^{(k)}_{\mathrm{SF}}=\frac{\left|\det{f_{\mathrm{A}}}\det{f_{\mathrm{B}}}\right|}{\left|4\widetilde{\mathrm{R}}^{AD}_{(00)}\tr\hspace{2pt}\left[(f^{\dagger}_{\mathrm{A}}f_{\mathrm{A}}\otimes f^{\dagger}_{\mathrm{B}}f_{\mathrm{B}})\rho^{\mathrm{AD}}\right]\right|}$, with $k=1,2$ take the values

 \begin{align}
&\Gamma^{(1)}_{\mathrm{SF}}=\frac{(\rho^{2}_{11}+\rho_{11}(\rho_{22}+\rho_{33})+\rho_{44}(\rho_{33}+\rho_{44})+\rho_{22}(2\rho_{33}+\rho_{44}))((\rho_{11}+\rho_{44})\sqrt{\rho_{22}\rho_{33}}-\sqrt{(\rho^{2}_{11}+\rho_{22}\rho_{33})(\rho^{2}_{44}+\rho_{22}\rho_{33})})}{2(\rho_{22}\rho_{33}-\rho_{11}\rho_{44})^{2}((\rho_{22}-\rho_{33})^{2}-(\rho_{11}-\rho_{44})^{2}-1)} ,\nonumber\\ 
& \Gamma^{(2)}_{\mathrm{SF}}=\frac{\rho^{2}_{22}+\rho_{22}\rho_{44}+\rho_{33}(\rho_{33}+\rho_{44})+\rho_{11}(\rho_{22}+\rho_{33}+2\rho_{44})}{2(1+(\rho_{22}-\rho_{33})^{2}-(\rho_{11}-\rho_{44})^{2})(\sqrt{\rho_{11}\rho_{44}}(\rho_{22}+\rho_{33})+\sqrt{(\rho^{2}_{22}+\rho_{11}\rho_{44})(\rho^{2}_{33}+\rho_{11}\rho_{44})})} ,
     \label{eq:caso5_fac2}
 \end{align}
here $\{\rho_{11},\rho_{22},\rho_{33},\rho_{44}\}$ are the diagonal elements of the X-form density matrix. $\Gamma^{(1)}_{\mathrm{SF}}$ corresponds to implement the projectors $\{\Phi^{\mathrm{BC}}_{0},\Phi^{\mathrm{BC}}_{1}\}$ and $\Gamma^{(2)}_{\mathrm{SF}}$ to implement  $\{\Phi^{\mathrm{BC}}_{2},\Phi^{\mathrm{BC}}_{3}\}$. Therefore the final obesity depends on the apply measurement.
\subsection{FS pathway}
\label{sec:appendix_5_B}

The {\it post-swapping-locally-filtered} state reads
\begin{align} \mathrm{R}^{\mathrm{AD}}_{\mathrm{FS}}=\frac{\widetilde{\mathrm{R}}^{\mathrm{AD}}}{\widetilde{\mathrm{R}}^{\mathrm{AD}}_{(00)}},
 \label{eq:FS_state}   
\end{align}
where $\widetilde{\mathrm{R}}^{\mathrm{AD}}=\frac{1}{4}(\mathrm{R}^{\mathrm{AB}}_{\mathrm{F}}\mathfrak{R}^{\mathrm{BC}}\mathrm{R}^{\mathrm{CD}}_{\mathrm{F}})$ according to \cref{r:general_swapping} but for locally-filtered states
\begin{align}
&\mathrm{R}^{\mathrm{AB}}_{\mathrm{F}}=\frac{\Lambda'_{A}\mathrm{R}^{\mathrm{AB}}\Lambda'^{T}_{B}\det{f'_{A}}\det{f'_{B}}}{\tr\hspace{2pt}\left[(f'^{\dagger}_{A}f'_{A}\otimes f'^{\dagger}_{B}f'_{B})\rho^{\mathrm{AB}}\right]},\ \ \text{and} \ \ 
    \mathrm{R}^{\mathrm{CD}}_{\mathrm{F}}=\frac{\Lambda''_{A}\mathrm{R}^{\mathrm{CD}}\Lambda''^{T}_{B}\det{f''_{A}}\det{f''_{B}}}{\tr\hspace{2pt}\left[(f''^{\dagger}_{A}f''_{A}\otimes f''^{\dagger}_{B}f''_{B})\rho^{\mathrm{CD}}\right]} .
    \label{eq:caso5_R3}
\end{align}
$f'_{\mathrm{A}}$,$f'_{\mathrm{B}}$ and  $f''_{\mathrm{A}}$,$f''_{\mathrm{B}}$ represents the optimal filters for $\rho^{\mathrm{AB}}$ and $\rho^{\mathrm{CD}}$, respectively. Quantum obesity of state in \eqref{eq:FS_state} reads
\begin{align}
\Omega^{\mathrm{AD}}_{\mathrm{FS}}=\frac{\left|\det{f'_{A}}\det{f'_{B}}\det{f''_{A}}\det{f''_{B}}\right| \Omega^{\mathrm{AB}}\Omega^{\mathrm{BC}}\Omega^{\mathrm{CD}}}{\left|4\widetilde{\mathrm{R}}^{AD}_{(00)}\tr\hspace{2pt}\left[(f'^{\dagger}_{A}f'_{A}\otimes f'^{\dagger}_{B}f'_{B})\rho^{\mathrm{AB}}\right]\tr\hspace{2pt}\left[(f''^{\dagger}_{A}f''_{A}\otimes f''^{\dagger}_{B}f''_{B})\rho^{\mathrm{CD}}\right]\right|} .
    \label{eq:caso5_OB3}
\end{align}
Considering states in XFS for the sources, as well as Bell measurements in part BC, quantum obesity reduces to
\begin{align}
\Omega^{\mathrm{AD}}_{\mathrm{FS}}=\Gamma_{\mathrm{FS}}\Omega^{2},
    \label{eq:caso5_OB4}
\end{align}
where $\Gamma_{\mathrm{FS}} = \frac{\left|\det{f'_{A}}\det{f'_{B}}\det{f''_{A}}\det{f''_{B}}\right|}{\left|4\widetilde{\mathrm{R}}^{AD}_{(00)}\tr\hspace{2pt}\left[(f'^{\dagger}_{A}f'_{A}\otimes f'^{\dagger}_{B}f'_{B})\rho^{\mathrm{AB}}\right]\tr\hspace{2pt}\left[(f''^{\dagger}_{A}f''_{A}\otimes f''^{\dagger}_{B}f''_{B})\rho^{\mathrm{CD}}\right]\right|}$ reads
\begin{align}
    \Gamma_{\mathrm{FS}}
    =
    \frac{
    1
    }{
    4
    \left(
    \sqrt{\rho_{11}\rho_{44}}
    +
    \sqrt{\rho_{22}\rho_{33}}
    \right)^{2}
    }.
    \label{eq:caso5_fac1}
\end{align}
$\Gamma_{\mathrm{FS}}$ corresponds to implement any of the projectors $\{\Phi^{\mathrm{BC}}_{0},\Phi^{\mathrm{BC}}_{1},\Phi^{\mathrm{BC}}_{2},\Phi^{\mathrm{BC}}_{3}\}$, therefore in this case the final obesity does no depend on the apply measurement.

\subsection{Reduction to the quantum obesity of almost-Bell-Diagonal states}
\label{sec:ABD}

A two-qubit X-state and its R-representation are given by:
{\small\begin{align*}
	\rho_X^{AB}=
	\left(\begin{array}{cccc}
		\rho_{11}&0&0&\rho_{14}\\
		0&\rho_{22}&\rho_{23}&0\\
		0&\rho_{23}^*&\rho_{33}&0\\
		\rho_{14}^*&0&0&\rho_{44}
	\end{array}\right)
 ,
	R_X
    =
    \hspace{-0.1cm}
    \left(
    \begin{array}{cccc}
    1&0&0& 2(\rho_{11}+\rho_{33})-1
    \\
    0&
    2
    (
    \text{Re}(\rho_{14})
    +
    \text{Re}(\rho_{23})
    ) 
    & 
    2(
    \text{Im}(\rho_{23})
    -
    \text{Im}(\rho_{14})
    )
    &0
    \\
    0&
    -2(
    \text{Im}(\rho_{14})
    +
    \text{Im}(\rho_{23})
    )
    & 
    2(
    \text{Re}(\rho_{23})
    -
    \text{Re}(\rho_{14})
    )
    &0
    \\
    2(\rho_{11}+\rho_{22})-1 &0&0& 1 - 2(\rho_{22}+\rho_{33})
    \end{array}
    \right)
    ,
\end{align*}}
with $\rho_{11}, \rho_{22}, \rho_{33}, \rho_{44} \in [0,1]$, $\rho_{14}, \rho_{23}\in \mathds{C}$ and $^*$ the complex conjugate. The constraint for almost-Bell-diagonal states can then be translated from the R-picture to the $\rho$-picture as $a_3=-b_3 \rightarrow \rho_{11}=\rho_{44}$ and $a_3=b_3 \rightarrow \rho_{22}=\rho_{33}$. We now analyse the quantum obesity coefficients for these two cases.

\subsubsection{Case 1 ($\rho_{11}=\rho_{44}$)}

Let us consider the case $\rho_{11}=\rho_{44}$, using the trace condition $\rho_{11}+\rho_{22}+\rho_{33}+\rho_{44}=1$ the obesity factors reduce to:
\begin{align}
&\Gamma_{SF}^{(1)}=\frac{1}{((\sqrt{\rho_{22}}-\sqrt{\rho_{33}})^{2}-1)^{2}}\\
&\Gamma_{SF}^{(2)}=\frac{1}{2((\rho_{22}+\rho_{33})-(\rho_{22}+\rho_{33})^{2})+\sqrt{((\rho_{22}+\rho_{33})^{2}+(\rho_{22}-1)^{2}+3\rho^{2}_{22}-2\rho_{33})((\rho_{22}+\rho_{33})^{2}+(\rho_{33}-1)^{2}+3\rho^{2}_{33}-2\rho_{22})}}\\
&\Gamma_{FS}=\frac{1}{((\sqrt{\rho_{22}}-\sqrt{\rho_{33}})^{2}-1)^{2}}
\end{align}
note that for the regime of allowed parameters $\rho_{22}+\rho_{33}\leq 1$ we have $\Gamma_{FS}/\Gamma_{SF}^{(1)}=1$. The case $\Gamma_{FS}/\Gamma_{SF}^{(2)}\geq 1$ is presented below considering the quantity $\Gamma_{FS}-\Gamma_{SF}^{(2)}$, which needs to be bigger than zero.
\begin{figure}[h!]
\centering
\includegraphics[width=0.35\columnwidth]{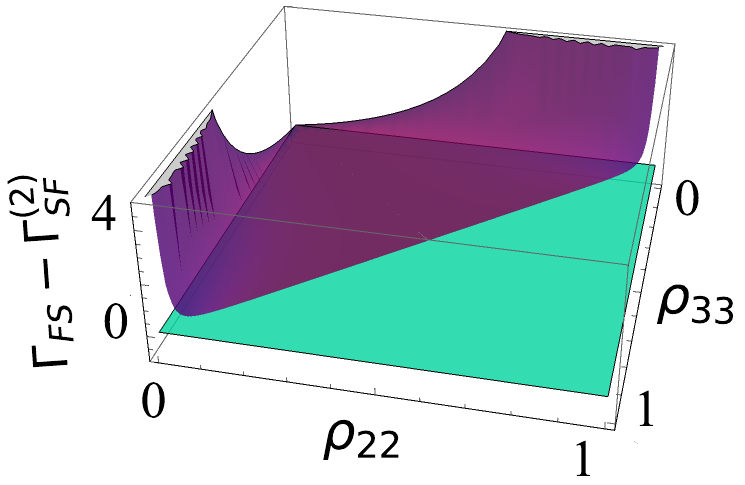}
\vspace{-0.2cm}
\caption{(Color online) 
The surface $\Gamma_{FS}-\Gamma_{SF}^{(2)}$ (purple) is always above the plane zero (cyan).
}
\label{fig:factors3D}
\end{figure}
\subsubsection{Case 2 ($\rho_{22}=\rho_{33}$)}
Now we consider the case $\rho_{22}=\rho_{33}$, the obesity factors reduce to:
\begin{align}
&\Gamma_{SF}^{(1)}=\frac{1}{2((\rho_{11}+\rho_{44})-(\rho_{11}+\rho_{44})^{2})+\sqrt{((\rho_{11}+\rho_{44})^{2}+(\rho_{11}-1)^{2}+3\rho^{2}_{11}-2\rho_{44})((\rho_{11}+\rho_{44})^{2}+(\rho_{44}-1)^{2}+3\rho^{2}_{44}-2\rho_{11})}}\\
&\Gamma_{SF}^{(2)}=\frac{1}{((\sqrt{\rho_{11}}-\sqrt{\rho_{44}})^{2}-1)^{2}}\\
&\Gamma_{FS}=\frac{1}{((\sqrt{\rho_{11}}-\sqrt{\rho_{44}})^{2}-1)^{2}}
\end{align}
similarly $\Gamma_{FS}/\Gamma_{SF}^{(k)}\geq 1$.
\twocolumngrid
\bibliographystyle{apsrev4-1}
\bibliography{bibliography.bib}

\end{document}